\documentstyle[psfig,astrobib]{mnv2}
\pssilent 

\def\ang{\mbox{\AA}}

\def\kms{\mbox{${\rm kms}^{-1}$}}

\def\deg{\hbox{$^\circ$}}

\def\asec{\hbox{$^{"}$}}

\def\msun{\mbox{${\rm M}_{\odot}$}}

\newcommand{\ale}{\ \raisebox{-.3ex}{$\stackrel{<}{\scriptstyle \sim}$}\ }
\newcommand{\age}{\ \raisebox{-.3ex}{$\stackrel{>}{\scriptstyle \sim}$}\ }
\def\etal   {et~al.\ }

\def\name#1 {{\it #1\/}}
\def\vol#1  {{\bf #1\/}}
\title[NGC 2865] {The shell elliptical NGC~2865: evolutionary
population synthesis of a kinematically distinct core}

\author[Hau et al.]  
{G. K. T. Hau$^{1,4}$, D. Carter$^{2}$, M. Balcells$^{3}$\\ 
$^1$Institute of Astronomy, Madingley Road, Cambridge CB3 0HA \\ 
$^2$Astrophysics Research Institute, Liverpool John
Moores University, Twelve Quays House, Egerton Wharf, Birkenhead, L41 1LD. \\ 
$^3$Instituto de Astrofisica de Canarias, C/ V\'\i a L\'actea, S/N, 
38200 La Laguna (Tenerife), Spain \\ 
$^4$Current address: Departamento de Astronom\'{\i}a y Astrof\'{\i}sica, Pontificia Universidad Cat\'{o}lica de Chile, Casilla 104, Santiago, Chile. \\
E-mail address: ghau@astro.puc.cl} 

\begin{document}

\def\zsun{\mbox{${\rm Z}_{\odot}$}}
\def\square{\vrule height 4.5pt width 4pt depth -0.5pt}
\def\threesquares{\square~\square~\square\ }
\def\remark#1{{\threesquares\tt#1~\threesquares}}

\newcommand{\mean}[1]{\mbox{$<$#1$>$}}
\newcommand{\h}[1]{\mbox{$h_#1$}}
\newcommand{\hbeta}{\mbox{H${\beta}$}}
\newcommand{\mg}[1]{\mbox{Mg$_#1$}}
\newcommand{\mgb}{\mbox{Mg$\ b$}}
\newcommand{\fe}[1]{\mbox{Fe$_{#1}$}}


\maketitle
\begin{abstract}

We report on the discovery of a rapidly co-rotating stellar and gas
component in the nucleus of the shell elliptical NGC~2865. The stellar
component extends $\sim 0.51 h_{100}^{-1}$ kpc along the major axis,
and shows depressed velocity dispersion and absorption line profiles
skewed in the opposite sense to the mean velocity.  Associated with it
is a young stellar population with enhanced \hbeta, lowered Mg and
same Fe indices relative to the underlying elliptical.  Its recent
star formation history is constrained by considering ``bulge+burst''
models under 4 physically motivated scenarios, using evolutionary
population synthesis.  Scenarios in which the nuclear component is
formed over a Hubble time or recently from continuous gas inflow are
ruled out.

A recent starburst can satisfy observational constraints {\it only if}
its population has metallicity 2.5--6.3 times that of the bulge.  The
nuclear iron to magnesium index ratio can be explained by a
temperature effect in the atmospheres of stars at main sequence
turnoffs between A3 and F4, during which the Fe indices of the burst
population are high enough to compensate for dilution effects. It is
therefore possible to modify line-index ratios (and hence the inferred
abundance ratio) simply by the presence of a young population {\it
with the same abundance}.  The high metallicity requirement suggests
self-enrichment, and burst duration longer than SN II feedback
timescale. No solution exists for bursts longer than $0.4$ Gyr. Burst
age estimates of $0.4$--$1.7$ Gyr are larger than that for the shells
(0.24 Gyr) assuming phase-wrapping.

No starburst is required if the nuclear component is composed of stars
with Fe abundance enhanced by $\sim 0.08$ dex relative to the
underlying elliptical, which are accreted by an event which truncated
the star formation. This relies on the abundance differences between
giant ellipticals and spirals.  The age-estimates of 0.1--0.4 Gyr in
this scenario are in closer agreement with those for phase-wrapped
shells.

Our results argue for a gas-rich accretion or merger origin for the
shells and kinematic subcomponent in NGC~2865. Arguments based on
stellar populations and gas dynamics suggest that one of the
progenitors is likely a Sb or Sc spiral. We demonstrate that despite
the age and metallicity degeneracy of the underlying elliptical, the
age and metallicity of the kinematic subcomponent can be
constrained. This work strengthens the link between KDCs and shells,
and demonstrates that a KDC can be formed from a late merger.
\end{abstract}

\begin{keywords}
galaxies: interactions -- galaxies: internal motions -- galaxies:
nuclei. -- galaxies: individual.
\end{keywords}

\section{Introduction} \label{sec:introduction} 

Observational and theoretical arguments suggest that shells in
elliptical galaxies and kinematically distinct cores (KDC)s may have a
common origin. Observationally, about 15--20\% of shell galaxies have
nuclear post starburst (E+A) spectra, suggesting that shell formation
process may involve some accretion or merger of gas-rich material with
associated nuclear star formation \shortcite{carter-etal88}.  In some
ellipticals the onset of the kinematic peculiarity is associated with
an enhancement in the metallic absorption-line strengths
(\citeNP{BS92}, \citeNP{carollo-danziger94}), which
may suggest that KDCs are products of early gas-rich mergers with an
associated metallicity enhancement.  The strongest association comes
from the finding by \citeN{Forbes92} that all of the 9 well
established KDCs and a further 4 out of the 6 ``possible KDCs''
possess shells.

The models proposed for the formation of shells and KDCs are very
similar, and many involve galaxy interactions or mergers. \citeN{Quinn84} considered mostly radial collisions of low-mass disks with a
spherical potential, and found that shells could be formed through the
process of phase wrapping. His ideas were further developed for
low-mass spheroidal companions \cite{DC86}, and for
non-radial orbits \cite{HQ88}. Hernquist and Quinn
further showed that mere mass-transfer during parabolic encounters
are capable of producing shells, demonstrating that a merger is not
necessary in the phase-wrapping model.

\citeN{HW92} extended the models in which a dwarf is
captured using a hybrid N-body/hydrodynamic code, tracking gas as well
as stars from the captured galaxy. They concluded that, while the
stars formed the shell structures, gas would settle into a compact
disc or ring in the centre of the primary. Shocks in the gas might
then trigger a nuclear starburst as seen in the spectra of
\shortciteN{carter-etal88}.

Mergers also provide the framework for most models of KDCs
(e.g. \citeNP{bender90rev}).  \citeN{fi88} noted that a
counter-rotating core such as that of IC~1459 could form in an
accretion of a gas rich satellite by gas-dynamical evolution and
ensuing star formation, or from the dynamical disruption of a
satellite near the core of the main galaxy, by simple stellar
dynamical processes.  \citeN{BQ90}, testing with
$N$-body models a scenario first proposed by \citeN{Kormendy84},
showed that KDCs including counter-rotating cores could be formed in
purely stellar dynamical simulations. They studied mergers between
elliptical galaxies of unequal mass, and found that the core
kinematics in the remnant depends mostly upon the orbital angular
momentum at a late stage of the merger, whereas the kinematics of the
outer regions are largely the original kinematics of the primary. Thus
in retrograde encounters a counter-rotating core can form.  When
counter-rotation forms via this mechanism, both primary and secondary
material counter-rotate at the core, and skewed line profiles are
produced \cite{Balcells92p221}.  Spiral-Spiral (SS) major mergers
provide promising mechanisms for the formation of KDCs.
\citeN{HB91} show an SS merger model with an embedded
counter-streaming gaseous disk.  Although the evolution of the gaseous
phase is still crude and star formation has not been included (hence
the predictive power of the model is limited), the physics is probably
on the right track.

\citeN{TW90} and \citeN{Thomson91} propose a
mechanism for the formation of shells in a weak (hyperbolic)
interaction between two massive galaxies. In this picture shells are
formed in a rotating ``thick disk'' component of an otherwise
non-rotating oblate elliptical, as an one-armed density wave excited
by the passage of another galaxy. When seen from different angles, the
shells are either interleaved or all-round, providing a natural
explanation for Type 1 (aligned) and Type 2 (all round) shells.
\citeN{HT94} further investigated whether a weak
interaction could explain the existence of KDCs, and proposed a
mechanism whereby a counter-rotating core could be formed by the
retrograde passage of a massive galaxy past a slowly rotating
elliptical with a pre-existing rapidly rotating central disk.  In this
picture it is the outer regions of the galaxy which are anomalous, the
direction of rotation is changed or reversed, whereas the central disk
is relatively unaffected.

A rather different model for the formation of shells was proposed by
\citeN{FNS80} and expanded by \citeN{Bertschinger85} and \citeN{WC86}.  In this picture shells
are formed as a result of star formation in shocked gas in the
interstellar medium of the galaxy, the shocks being caused by outflow
from the nucleus. With the measurement of shell colours, which are not
much bluer than the main body of the galaxy, the failure to detect
either ionized or neutral gas associated with the shells except in a
very few cases, and the discovery of the interleaving of shells on
opposite sides of the nucleus in many shell galaxies, this picture has
fallen into disfavor, though \citeN{lowenstein-etal87} reconcile this
model with the last of these observations. 

Motivated by the observational and theoretical arguments outlined
above, we initiated a programme to look for KDCs in a sample of shell
galaxies drawn from the \citeN{MC83} catalogue, to
investigate whether there is a common formation mechanism for shells
and KDCs.  In this paper we report the results for NGC~2865.

\begin{figure}
  \centerline{ \hbox{
   \psfig{figure=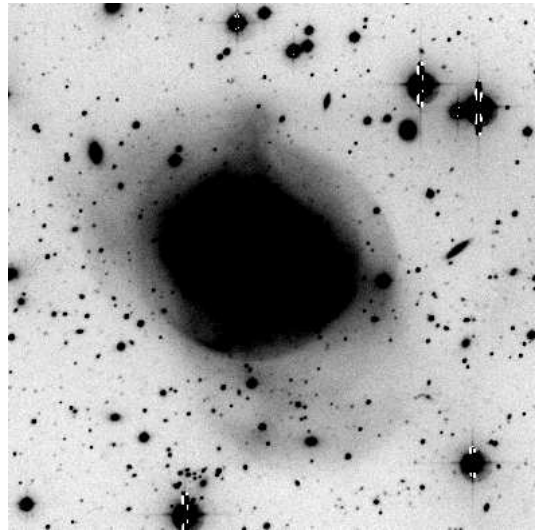,width=70mm}}}
\caption[]{R-band image of NGC~2865, showing the bright, chaotic shells.}
\label{fig:n2865small}
\end{figure}

NGC~2865 is listed in the \citeN{MC83} catalogue of shell
galaxies.  It is classified as E3+ in RC3 \cite{rc3}, and E4 in RSA
\cite{rsa}.  Taking $m_b=12.35$ from RC3, $A_B=0.27$, $h_{100} =1$ and
$cz=2530\,\kms$, its $M_B$ of -20.02 places it on the brighter side of
the dividing line between low-luminosity and giant ellipticals
($M_B^{UB} = -18.99$ with $h_{100} =1$; \citeNP{davies-etal83}). CCD
photometry reveals surface brightness that is consistent with a
$r^{\frac{1}{4}}$ law out to $\sim 60 \asec$ \cite{RBS94}. Its chaotic shells are bluer than the main body of the galaxy
at $\sim$84-98\% confidence level, and contribute $\sim 11\%$ -- $
22\%$ of the total luminosity \cite{fort-etal86}. Fort \etal noted the
colour indices of the shells resemble those of an Sb or Sc galaxy. The
galaxy appears disturbed in a deep R-band image, and in the outer
parts the shells are brighter than the main body
(Fig.~\ref{fig:n2865small}). The mild ``E+A'' spectrum
\cite{carter-etal88} in the nucleus suggests some recent star
formation. Carter \etal estimated that light from A stars contributed
$30\pm10\%$ of the light at the nucleus.  NGC~2865 is detected by IRAS
at $100\ \rm{Jy}$, with an estimated dust mass of $0.17 \times 10^6
\msun$ \cite{roberts-etal91}. These properties are evidence for a
recent merger or accretion that involves a gas-rich disk galaxy.

By performing population synthesis using galaxies and galactic
clusters as templates, \citeN{BA87} find that a starburst
has occurred, and derived an age of $1.2\pm0.3$ Gyr for it.

\shortciteN{schiminovich-etal95} observed NGC~2865 with the VLA, and
found $6 \times 10^8h_{100}^{-2}\msun$ of cold gas. Surprisingly, the
gas is found in a broken ring in the outer regions of NGC~2865, and is
displaced to the {\it outside} of the shells and the loops. The HI
velocity field is smooth, and has the same sense and magnitude as the
inner stars. Fits of rotation curves to the intensity-weighted
velocity field yielded a circular velocity of $240 \pm 15\,\kms$ and
an inclination of $65 \deg \pm 5\deg$, at a position angle of $310
\deg \pm 5\deg$, close to the photometric major axis.  Schiminovich
\etal derived a mass-to-light ratio $M/L_B$ of $33 \pm 4 h_{100}$ at
large radii, which is about 5 times greater than $M/L_B$ at the core.

There is a high degree of rotational support for NGC~2865.  Stellar
kinematics along 4 axes previously measured by \citeN{Bettoni92}
shows $v/\sigma \sim 1.3$, greater than required for an isotropic
rotator of the same flattening.  Bettoni further argued that NGC~2865
must be oblate, because the velocity gradients along the 4 position
angles are consistent with a sinusoidal function of position angle,
with the fastest rotation along the major axis. Peculiar kinematics
were not found in the Bettoni data.

Whilst both \citeN{Bettoni92} and \shortciteN{schiminovich-etal95}
point to a merger as responsible for the shell formation, the physical
pictures favored are different. Citing the regular stellar velocity
field and lack of emission lines, Bettoni concludes that his results
are consistent with the \citeN{DC86} model in which a
small satellite is absorbed and disrupted by an almost oblate E3.5
galaxy, and that the collision has only marginally perturbed the
underlying potential. \shortciteN{schiminovich-etal95}, however, favor
the equal-mass spiral-spiral merger scenario (\citeNP{HS92}, \citeNP{HM95}) in which the shells and
other fine features are formed out of material raining back into the
core after the collision, citing as support the near spiral $M_{{\rm
HI}}/L_{{\rm B}}$ in the outer regions and the close matching between
the stellar and gas kinematics.  They concluded, however, that the
displacement of the gas outward from the stars is not supported by any
of the current merger scenarios.

In this paper we report on the discovery of a kinematically distinct
core in the nucleus of NGC~2865. We recover the line-of-sight velocity
distribution along two axes, and find a small, rapidly rotating
stellar component with associated ionised gas. We measure the
abundance distribution of NGC~2865, and find that a young stellar
population is associated with the KDC. A crude modelling of the
dilution effects using stellar templates shows that major star
formation must have stopped at least $\sim 0.5$ Gyr ago. The KDC shows
a iron to magnesium index ratio higher than the light from the
underlying galaxy.  By considering the evolution of a post-starburst
spectrum with time, we find that this may be explained as a
temperature effect in the atmospheres of stars at main-sequence
turn-offs between A3 and F4, instead of a change in abundance
ratio. We then constrain the star-formation history of the KDC, aided
by the spectral evolution library of Bruzual \& Charlot (1997). We
derive constraints on physical parameters, and find that the KDC was
formed either by a metallicity-enhanced starburst, or by accretion of
a spiral into a giant elliptical with a different abundance ratio,
during which star formation is truncated.

The layout of this paper is as follows: in \S 2 the observation and
data reduction procedures are described. Stellar kinematics derived
from the absorption-line profile, and gas kinematics derived from the
[OIII] emission line are presented in \S 3. In \S 4 the radial
gradients of absorption line indices are presented. In \S 5 \& 6 we
put initial constraints on the age of the starburst by considering the
effect of dilution by a young stellar population. In \S 7, the star
formation history is further constrained using evolutionary stellar
population synthesis and considering a ``bulge+burst'' model under 4
scenarios of star formation. This is followed by a discussion in \S
8. In order not to distract from the main scientific thrust, the
reliability of the methods employed in extracting the absorption-line
profile is investigated by Monte Carlo simulations in
Appendix~\ref{sec:noisetest}, whilst the dependence on stellar
templates is investigated in Appendix~\ref{sec:template-tests}.

\section{Observations} \label{sec:observations}

\subsection{The data} 

\begin{table}
\centering
\caption
{Instrumental setup for the observations.}
\label{tab:instruments}
\vspace{0.5cm} 
\begin{tabular}{ll}
\hline
Date              & April 1995 \\
Spectrograph      & RGO        \\
Camera            & 25 cm \\
Grating (lines/mm)& 1200V      \\
Detector          & TEK 1024 CCD  \\
Gain (e$^{-}$/ADU)          & 2.7 \\
Readout noise (e$^{-}$/pix) & 4.8 \\
Scale along slit (\asec/pix)          & 0.77 \\
Dispersion (\ang/pix)       & 0.79 \\
Velocity per bin (kms${\rm ^{-1}})$ & 42.7 \\
Wavelength range (\ang)     & 4850--5610 \\
Instr FWHM ($\ang$) &  2.1  \\
\hline
\end{tabular}
\end{table}

Long-slit spectra of NGC~2865 were taken with the RGO spectrograph on
the Anglo Australian Telescope in April 1995.  The instrumental setup
is summarized in Table~\ref{tab:instruments}.  The spectral and
spatial resolution are significant improvements over earlier studies
of this galaxy \cite{Bettoni92}. The spectra span the wavelength
range from $4850\ang$ to $5610\ang$, covering the \hbeta, Mg \& Fe
absorption lines as well as the [OIII] emission line at $\sim 5007
\ang$. With a $2.6\asec$ slit, the exposures were $3600$ seconds along
the major axis (P.A. $153\deg$) and minor axis (P.A. $63\deg$). The
seeing was $0.8 \asec$ FWHM for the major axis data and $0.9 \asec$
for the minor axis data.

The data were bias subtracted, flat-fielded, cleaned of cosmic rays
and wavelength calibrated using the standard procedures in the AAO
FIGARO data reduction package. The spectra were rebinned to a
logarithmic wavelength scale to allow cross-correlation and
Gauss-Hermite analysis.

In \S~\ref{sec:mg} we find that the central Mg indices are depressed,
but not the Fe. This may arise if there exists a focusing problem with
the spectralgraph under good seeing conditions, first mentioned by
\citeN{DSP93}, and further discussed by
\citeN{GJ93} and \citeN{MSBW98}.  This arises as the
continuum bandpasses of \mg1 and \mg2 has a wide wavelength
separation, and an out-of-focus camera will distribute the fluxes at
the end of the central spectra away, thus artifically lowering the
continuum level and in turn the absorption-line index.  Such a problem
is not present in our data.  The CCD frames are well and uniformally
focused, with no detectable spatial variation in focus when an arc
lamp exposure is examined. 

Template spectra of 3 stars HD~80170 (K5III), HD~107328 (K0IIIb) and
HD~109379 (G5IIb) were taken with the same instrumental setup as the
galaxy observations. After their radial velocities measured by
cross-correlation, the template spectra were de-redshifted to the
reference frame of HD~109379, which has a heliocentric velocity of
$-7.8\,\kms$ \cite{MM85p299}. This value is added to the
measured velocities in subsequent analysis.

\subsection{Gauss-Hermite parameterization of LOSVD}\label{sec:technique}

Following \citeN{vdmf93}, the line-of-sight velocity distributions
(LOSVDs) $L(w)$ are taken to be Gaussians with small Hermite
deviations:
\begin{equation} 
        L(w) = \gamma \frac{\alpha (w)}{\sigma} \left[
          1+\sum_{n=3}^{N} h_n \ H_n(w)\ \right]
\label{eq:losvdgausshermit}
\end{equation} 
where
\begin{eqnarray} 
      w         &=& \frac{v-V}{\sigma} \\
      \alpha(w) &=& \frac{1}{\sqrt{2 \pi}}e^{-\frac{w^2}{2}}
\end{eqnarray} 
and $\gamma$, $V$, $\sigma$ are the line-strength, mean velocity,
velocity dispersion respectively. $H_n$ is a Hermite polynomial of
order $n$ and with a corresponding amplitude $h_n$. Expressed
explicitly, $H_3$ and $H_4$ are:
\begin{eqnarray}
      H_3(w) &=& \frac{1}{\sqrt{3}} ( 2 w^2 -1 ) \\ H_4(w) &=&
      \frac{1}{\sqrt{6}} ( 2 w^3 -3w )
\end{eqnarray} 
Thus the LOSVD is parameterized by the variables ($\gamma$, $V$,
$\sigma$, $h_3,...,h_N$), and its shape is a perfect Gaussian when the
$h_n$ are zero. The Gauss-Hermite moments \h3 and \h4 are meaningful
parameters as they are related to the skewness and the kurtosis of the
distribution respectively, and are useful indicators of unusual core
kinematics. A large \h3 with opposite sign to the mean velocity is
typical of rapidly-rotating systems, whilst \h4 is related to the
velocity dispersion anisotropy in the galaxy core
\cite{vdmf93}. Although the higher moments are more difficult to
interpret, they are still useful for identifying structures in the
residuals of the model fits, and furthermore a large signal in them
indicates the breakdown of the near-Gaussian approximation, therefore
parameters up to \h6 are measured and plotted.

In the rest of this section, the method of extracting the LOSVD is
described. Readers not interested in this should go straight to
\S~\ref{sec:kinematics}. To recover the LOSVD, we use the program {\tt
kinematics}, written and kindly made available to us by
H.-W. Rix. This program implements the parameterization of
\citeN{vdmf93} into the pixel space fitting routines of \citeN{RW92}; see also \shortciteN{vdm-etal94}.  First a stellar
template is convolved with an initial guess of $L(w)$ to give a model
galaxy spectrum. The parameters ($\gamma$, $V$, $\sigma$) are
recovered when the noise weighted $\chi^2$ of the model and the galaxy
spectrum are minimized. Each $h_n$ is estimated by keeping all lower
moments fixed in the $\chi^2$ fitting, and this process repeated until
$n=6$.
\footnote{In this paper we give ($v$, $\sigma$) as those defining the
best fitting $L(w)$ according to equation~\ref{eq:losvdgausshermit},
which are slightly different to ($V$, $\sigma$) defined as moments of
$V$ and $V^2$ of a LOSVD, as pointed out by \citeN{vdmf93}, who also
give expressions of the formulae for conversion.}  The errorbar of
each parameter is taken as the $63\%$ confidence region, assuming all
other parameters are fixed.

The wavelength range 4910--5503 $\ang$ is well known to give template
mismatch problems.  To minimize this, an optimal stellar template is
employed, taken as the linear combination of the 3 templates that
minimizes the $\chi^2$ of a Gaussian profile, after first estimating
($\gamma$, $V$,~$\sigma$).  The optimal template, derived separately
for each point along the slit, is used to recover the LOSVD, modelled
up to \h6. The galaxy nucleus is located by a Gaussian fit to the
light profile of the central few pixels. Due to the presence of a
foreground star, no results are presented beyond $-6\arcsec$ of the
nucleus along the minor axis.

Experimentation shows that summation to continuum $S/N \age 40$ per
pixel gives satisfactory results.  The sensitivity of the measured
parameters to noise and template mismatch are investigated in
Appendices \ref{sec:noisetest} \& \ref{sec:template-tests}. The
conclusions from Monte Carlo simulations in Appendix
\ref{sec:noisetest} are the following: Hermite moments up to \h6 can
be recovered to within 0.01--0.02 up to $3 \asec$ from the nucleus. A
small but significant displacement in a parameter is introduced if the
Hermite moment two orders up is large.  The parameter least sensitive
to noise is $v$, whilst $\sigma$ can be affected by $S/N$ and by power
in \h4, with a signal of $0.05$ in \h4 causing a shift of $\sim
+4\,\kms$.  There are no problems associated with imperfect sky
subtraction, even when 12 spectra are combined to achieve the desired
$S/N$. Therefore along the major axis the LOSVD is modelled up to \h6
for $|r| \ale 3 \asec$ and up to \h5 for $|r| \ale 6 \asec$, and to a
smaller extent along the minor axis due to the lower $S/N$.

The conclusions from Appendix~\ref{sec:template-tests} are the
following: Template mismatch cannot explain the radial profile of the
parameters (apart from possibly \h4).  The maximum offsets are of
order $6\,\kms$ in $v$ \& $\sigma$, and $0.026$, $0.026$, $0.020$ \&
$0.018$ in \h3, \h4, \h5 \& \h6 respectively. The observed zero-point
offsets in \h3, \h5 \& \h6 may be attributed to template mismatch, but
the positive offset in \h4 is marginally significant even in the
extreme case. Thus we think the positive \h4 reported below has a
physical origin.

\subsubsection{Regions excluded from direct fitting} \label{sec:emission}

Emission lines can potentially introduce systematics in the absorption
line profile analysis, and further affects the measurement of
absorption-line indices if they fell within one of the bandpasses.
Hidden [NI] emission at $5199\ang$ can artificially lower the \mgb\
equivalent width by $0.4$--$2\ang$ \cite{GE96},
but given that no significant residuals can be seen at $5199 \ang$
from the $\chi^2$ fits, it is probably negligible. From the $\chi^2$
fits we find some extremely weak residuals within the continuum
bandpasses of \mgb\ and \fe{5270} coincide with the location expected
of some [Fe] or [CrI] emission lines. Whilst these residuals may be
caused by template mismatch, nevertheless the regions
$4951.0$--$4955.9\ang$, $5141.27$--$5158.1\ang$ \&
$5519.6$--$5533.8\ang$ (in the galaxy's rest frame) are excluded from
the line-profile analysis. The \hbeta\ absorption line, [OIII]
($5002.7$--$5014.1\ang$) and [NI] ($5196.4$--$5201.5\ang$) emission
lines are also excluded, as well as occasional regions affected by
cosmic rays. Experimentation shows that excluding a few pixels from
the $\chi^2$ fits does not affect the recovery of the kinematic
parameters because information is derived from the entire spectrum.

\section{Core kinematics of NGC~2865} 

\begin{figure*}
  \centerline{ \hbox{
   \psfig{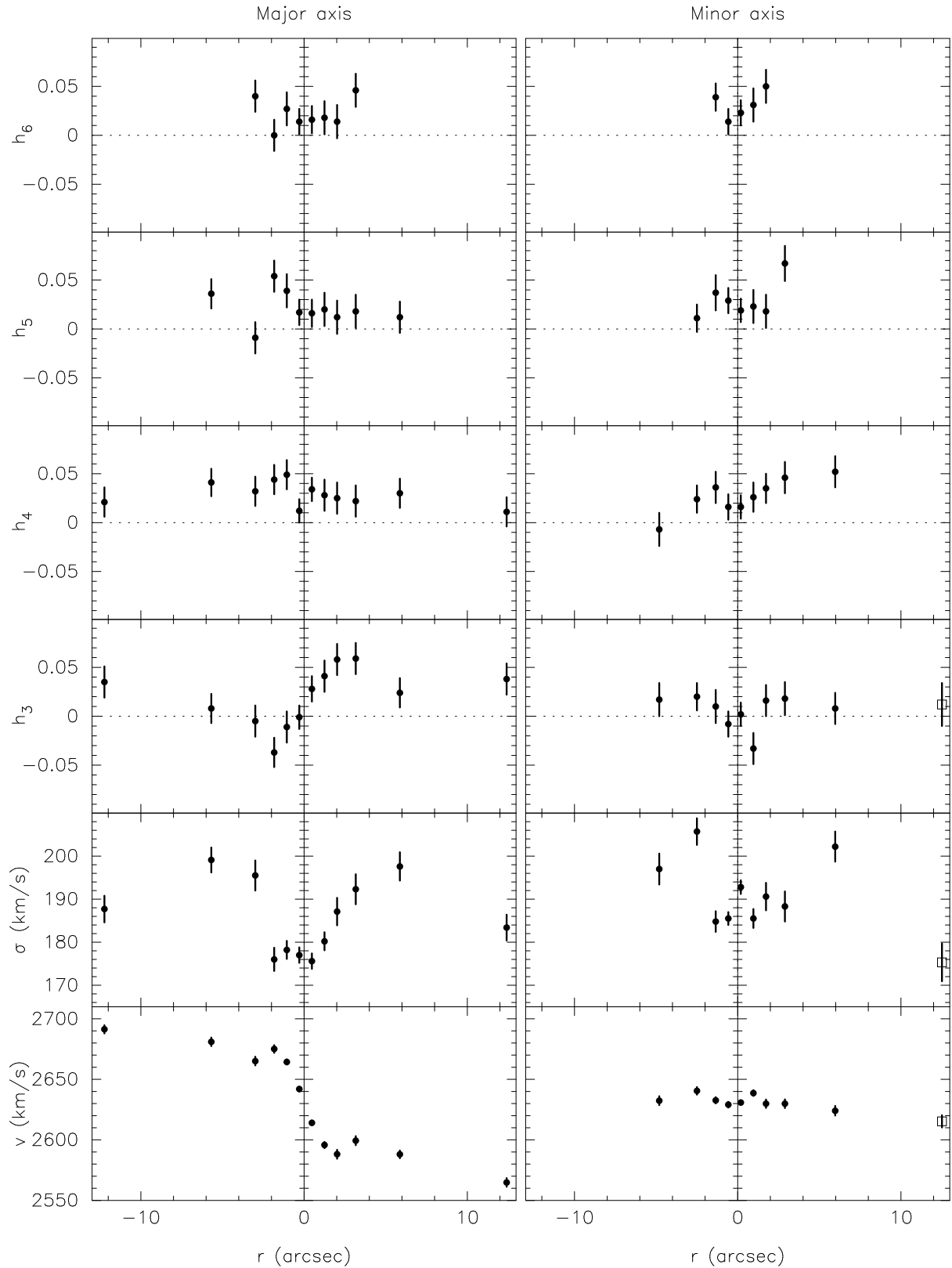}}}
\caption[]{The $v,\sigma,h_3,h_4,h_5$ \& $h_6$ along the major and
minor axes of NGC 2865. The results with solid dots are from spectra
with $S/N \age 40$ per pixel, whilst the point plotted with an open
symbol has $S/N \approx 30$. The $1 \sigma$ errorbars are also
plotted.}
\label{fig:n2865hermplot}
\end{figure*}

\subsection{Kinematics of stars derived from absorption line profiles}
\label{sec:kinematics}

The results of the absorption line profile analysis are presented in
Fig.~\ref{fig:n2865hermplot}.  Along the major-axis the rotation curve
is symmetric, and is near solid-body for $|r| \age 3 \asec$, reaching
$\sim \pm 63\,\kms$ at $\sim 12 \asec$. The velocity dispersion
increases inward from $\sim 186\,\kms$ at $|r| \approx 12\asec$ to a
maximum of $\sim 200\,\kms$ at $|r| \approx 6\asec$ . The \h3 is
consistent with $\sim 0.02$ beyond $\sim 4\asec$. Within $3 \asec$ of
the nucleus, the rotation is enhanced relative to the general trend,
with associated $\sigma$ that is depressed by $\sim 23\,\kms$, and
point-symmetric \h3\ with magnitude up to 0.05.  The latter indicates
the line profiles are skewed in the opposite sense to the mean
velocity.  These results suggest that NGC~2865 harbours a cold,
rotating nuclear stellar component that is co-rotating with the
underlying elliptical, giving rise to the steeper velocity gradient
and point-symmetric \h3 in the inner $3 \asec$. This component has
high surface brightness and the light contributed by it lowers the
central velocity dispersion. Along the minor-axis the rotation has an
upper limit of $15\,\kms$, and there is no significant signal in \h3
except a zero-point of +0.01.  The lack of rotation along the minor
axis is consistent with the kinematic profiles of \citeN{Bettoni92}, and suggests that the intrinsic shape of NGC~2865 is close to
oblate.

The minor axis $\sigma$ within $3 \asec$ of the centre are $\sim 9\
\kms$ higher than those in the same region along the major axis. The
region of lowered $\sigma$ is slightly more extended and shallower,
and the depression of $\sim 18\,\kms$ is smaller than the major axis.
This difference could be due to the different pixel geometry along the
2 axes. For the major axis data each pixel samples $0.7 \asec$ ($0.09
h_{100}^{-1}$ kpc) along the major axis and $2.6 \asec$ ($0.33
h_{100}^{-1}$ kpc) along the minor axis, and {\it vice versa} for the
minor axis data. If the kinematic subcomponent is a rapidly rotating
disk aligned along the major axis and its axis inclined at an angle
$i$ with respect to the line-of-sight, then the major axis pixels
sample mainly the coherent rotation of the stars, with the dispersion
mainly from the $\sigma_z {\rm cos}~i $ component. In comparison, the
minor axis pixels are 3.7 times wider along the disk and $0.27$ times
narrower perpendicular to it, hence a major contribution to the
nuclear $\sigma$ is from the velocity gradient over $0.33
h_{100}^{-1}$ kpc along the disk, and the contribution from $\sigma_z
{\rm cos} i $ is smaller. Thus if the disk is cold and rotating
rapidly it is possible to measure a different nuclear $\sigma$ along
the major and minor axes.

Along both axes, the line profile is significantly Lorentzian with
$\h4$ up to $0.05$. There is a small offset of $\sim 0.02$ in \h5 and
\h6 along both axes, albeit these 2 parameters are more sensitive to
noise and template mismatch.  Appendix~\ref{sec:noisetest} shows that
a significant signal in \h6 of $\sim 0.05$ may artificially enhance
\h4 by $\sim 0.02$, whereas an \h4 of $0.05$ may artificially shift
$\sigma$ upwards by about $4\,\kms$.  Therefore if the observed \h4
and \h6 have a physical origin, the true \h4 and $\sigma$ may be lower
than measured by about $0.01$ and $4\,\kms$ respectively.

The central positive \h4 is one of the outstanding properties of the
nuclear kinematics of NGC~2865.  Our error estimates (Appendix A \& B)
indicate that the signature is real, unless there is an extreme
template mismatch, unlikely given the good match in metallicity
between the templates and the galaxy. Positive \h4\ is normally
attributed to predominance of radial orbits. However, radial orbits
yield an outwardly decreasing $\sigma$ profile for a constant $M/L$,
which is not seen at the nuclear region.  We find a similarly positive
\h4\ signature in another shell lenticular NGC~474 \cite{hau96iau}.
Such positive \h4\ values are rare in the nuclei of ellipticals
(e.g. \citeNP{BSG94}), and could contain
information on the type of merger which produced the shells.

The central velocity dispersion is about $200\,\kms$ without the
KDC. Taking the rotation velocity of $\sim 240\,\kms$ at the outermost
points of Fig.1b in \cite{Bettoni92}, we arrive at a $v/\sigma \sim
1.2$, slightly smaller than the value of 1.3 estimated by Bettoni
1992. This difference can be attributed to the central dispersion
value of $180\,\kms$ adopted by Bettoni, who did not detect the KDC.
We measure a mean ellipticity of $\sim 0.26$ for $r \age 30
\asec$. From Fig. 1 of \citeN{Binney78}, we find that the
$v/\sigma$ of NGC~2865 is twice that expected for an oblate isotropic
spheroid of the same flattening.

\subsection{Kinematics of ionized gas} \label{sec:gas}

\begin{figure*}
  \centerline{ \hbox{\psfig{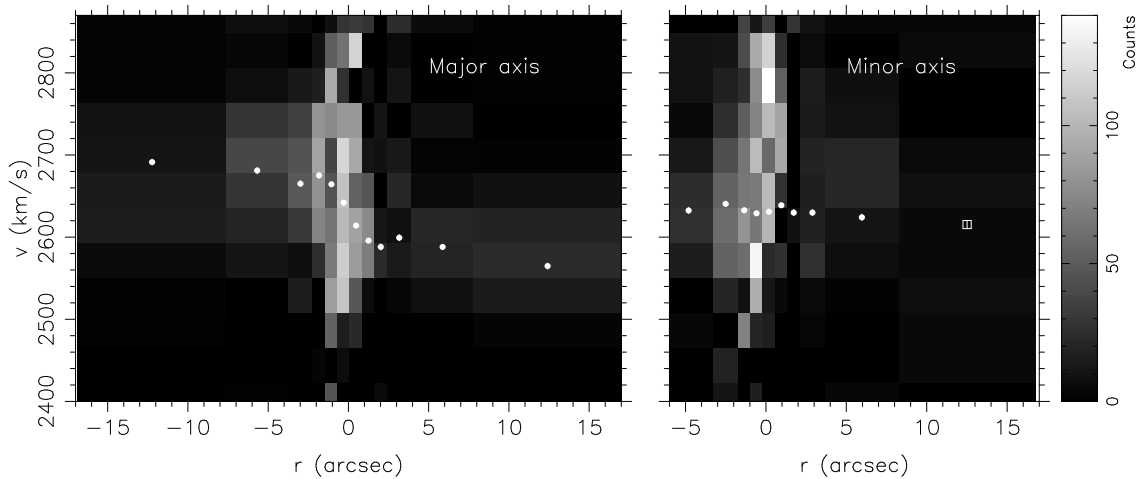}} }
\caption[]{The gas kinematics from the [OIII] emission line at
$5007\ang$ along major and minor axes of NGC 2865. The stellar
velocity curves from Fig.~\ref{fig:n2865hermplot} are over-plotted in
white. No data is plotted for $r < -6\asec$ on the minor axis due to
light contamination by a foreground star.}
\label{fig:n2865o3}
\end{figure*}

In this section we present the line-of-sight velocity distribution of
the ionized gas inferred by the profile of the weak [OIII] emission
line, obtained directly from the residual map of the Gauss-Hermite
analysis.  We find this method very successful and more objective than
fitting the baseline by hand. The results for the 2 axes are plotted
in Fig.~\ref{fig:n2865o3}, with the stellar rotation curve from the
previous section over-plotted.  Because the [OIII] emission is weak,
we feel that Fig.~\ref{fig:n2865o3} allows only a qualitative
description. At the nucleus, light from the underlying galaxy
contributes up to $\sim 4700$ counts per pixel, therefore the $S/N$
per pixel is about 3, explaining the noisier appearance near the
nucleus.

Along the major axis, the gas co-rotates with the stars. At the
location where the kinematically distinct component is observed in the
stellar component, the [OIII] line appears to be split into a
parallelogram shape with a peak-to-peak velocity of $\sim 260\,\kms$.
Along the minor axis there is no observable rotation.  Within $1\asec$
of the nucleus, the [OIII] line appears to have 2 peaks--- at $\sim
2570\,\kms$ and $\sim 2780\,\kms$, but we note that the data along the
minor axis have a lower $S/N$. As a rough estimate, assuming the gas
is in circular orbits which are inclined at $45\deg$
(\S~\ref{sec:hbeta}), and taking a projected rotational velocity of
$130\,\kms$ along the major axis, the total mass within $1.9 \asec$
($0.24\ h_{100}^{-1}$ kpc) of the nucleus is estimated to be $1.9
\times 10^9 h^{-1}\msun$.

Despite the low $S/N$, the major and minor axes [OIII] brightness are
well fit by power-law profiles, and not an exponential profile. The
power-law indices for the major and minor axes profiles are $0.651$
and $0.891$ respectively, indicating the light profile is steeper along
the minor axis, and that the flattening along the major axis increases
with distance from the nucleus. 

The results from this section show that the stars and gas rotate with
approximately the same magnitude and direction about the minor
axis. The flattening and lack of rotation along the minor axis
suggests that the gas is a co-rotating disk associated with the
nuclear kinematic sub-component.

\section{Absorption line indices} \label{sec:indices}

\begin{figure*}
  \centerline{ \hbox{
   \psfig{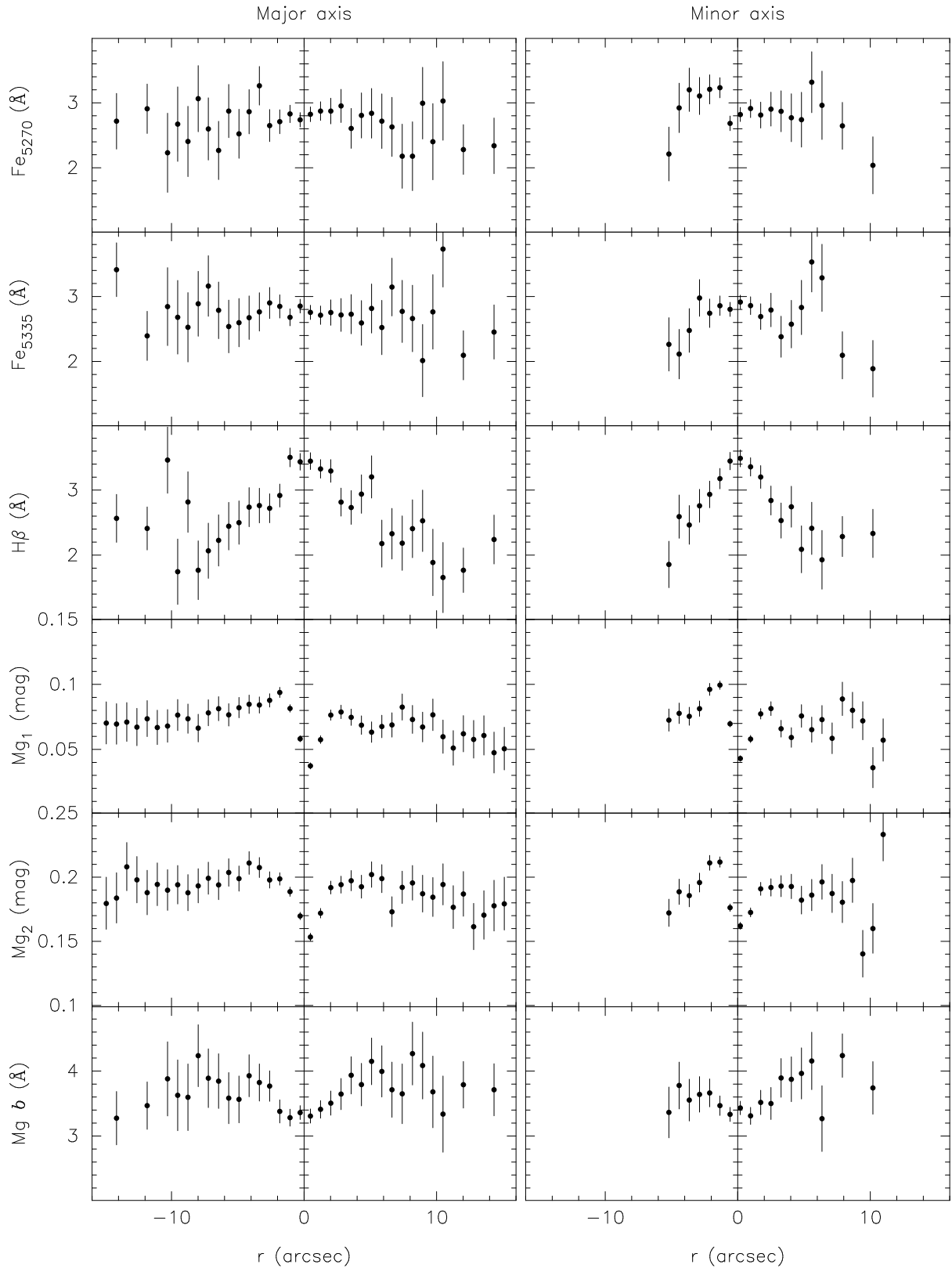}}}
\caption[]{The radial profiles of the absorption line indices
\fe{5270}, \fe{5335}, \hbeta, \mg1, \mg2 and \mgb\ along the major and
minor axis of NGC 2865. }
\label{fig:n2865indices}
\end{figure*}

\subsection{Procedures} \label{sec:procedures}

\begin{table*}
\centering
\caption{The bandpass definitions of measured absorption line indices, according to Worthey et al. (1994).}
\label{tab:line-intervals}
\vspace{0.5cm} 
\begin{tabular}{llll}
\hline
Feature & Type & Central Bandpass (\ang) & Continuum Bandpasses (\ang) \\
\\

\hbeta &Atomic absorption line & 4847.875--4876.625  & 4827.875--4847.875, 4876.625--4891.625 \\
\mg1  & Molecular band           & 5069.125--5134.125  & 4895.125--4957.625, 5301.125--5366.125 \\
\mg2  & Molecular band           & 5154.125--5196.625  & 4895.125--4957.625, 5301.125--5366.125 \\
\mgb &Atomic absorption line & 5160.125--5192.625  & 5142.625--5161.375, 5191.375--5206.375 \\
\fe{5270} &Atomic absorption line     & 5245.650--5285.650  & 5233.150--5248.150, 5285.650--5318.150 \\
\fe{5335} &Atomic absorption line     & 5312.125--5352.125  & 5304.625--5315.875, 5353.375--5363.375 \\
\hline
\end{tabular}
\end{table*}

\nocite{worthey-etal94apjs94}

In this section the radial profiles of the absorption line indices are
presented. Line indices are extracted from individual, de-redshifted
spectra following the recipe of \shortciteN{ffbg85}, according to the
bandpass definitions of \shortciteN{worthey-etal94apjs94} adopted in
the Bruzual \& Charlot models (\S~\ref{sec:bcmodelling}). Passbands
for the various indices and continua are given in
table~\ref{tab:line-intervals}.  The atomic absorption lines \hbeta,
\mgb, \fe{5270} and \fe{5335} are measured as equivalent widths in
Angstroms, whilst the molecular bands \mg1 \& \mg2 are measured in
magnitudes. The indices are corrected to a zero dispersion system by
applying correction factors, estimated by broadening the 3 stellar
templates to dispersions ranging from $160\,\kms$ to $210\,\kms$. The
correction factor for each $\sigma$ is taken as the mean of the
correction estimated for the 3 stars.  The typical correction factor
at $\sigma=200\,\kms$ for \mg1, \mg2, \mgb, \hbeta, \fe{5270} and
\fe{5335}, are $0.976$.  $0.994$, $0.873$, $0.968$, $0.886$ \& $0.865$
respectively (where the measured line indices should be divided by
these values).  The spread in these values $\sigma_C$ are $0.006$,
$0.005$, $0.009$, $0.028$, $0.012$ and $0.009$
respectively. \footnote{ The dispersion correction factors for \hbeta\
are estimated by redefining the central bandpass to $4861$--$4877
\ang$ and using only the red continuum level, because at zero redshift
the blue continuum bandpass of the template stars is shifted out of
the observing window.  This method accounts for a larger $\sigma_C$
compared to other line indices, but is not a problem as the error in
\hbeta\ is dominated by photon shot noise.}  The quoted errors
$\sigma_{\rm I}$ in the measured line indices take into account both
the error due to photon shot-noise and $\sigma_{\rm C}$. The indices
are not degraded to the Lick spectral resolution of $\sim 8 \ang$
FWHM, and because our data are not flux calibrated, and no stars with
Lick indices are observed, there is some uncertainty in the zero
points.  This does not affect the interpretation of our results;
firstly because the \mg1 and \mg2 indices (which show radial
variation) are extremely insensitive to line-broadening, and secondly
because the difference in each index (e.g. $\Delta \hbeta$, $\Delta
\fe{5335}$ \& $\Delta \mean{Mg}=(\mg1+\mg2)/2$) is the free parameter
in the models.  Furthermore, within $8\asec$ of the nucleus along the
major axis, the average \mg2 of $0.192 \pm 0.014$ mag agrees with that
measured by the Lick group ($0.208 \pm 0.02$ mag;
\citeN{trager-etal97}). \footnote{A proper comparison would require
convolving the window function adopted by the Lick group with our
data.}  The average \fe{5270} and \fe{5335} of $2.75 \pm 0.24\ \ang$
and $2.77 \pm 0.17\ \ang$ also agree within error with the Lick group
values ($3.12 \pm 0.4\ \ang$ and $2.82 \pm 0.4\ \ang$). Thus we are
confident that our measurements are close to the Lick scale, and the
zero-point offsets are the same order of magnitude as the smallest
errorbars.

\subsection{Radial profiles of absorption-line strengths}

\subsubsection{\hbeta} \label{sec:hbeta}

The radial profiles for the \fe{5270}, \fe{5335}, \hbeta\, \mg1, \mg2
\& \mgb\ indices are presented in Fig.~\ref{fig:n2865indices}.  The
\hbeta\ is $\sim 2\ \ang$ at large radii, and increases gradually
towards the nucleus, where it reaches $\sim 3.5\ \ang$.  \hbeta\
absorption is strongest in A-type stars and a large equivalent width
indicates the presence of a young stellar population. The major axis
\hbeta\ profile indicates that the fraction of the young population
along the line-of-sight gradually increases towards the nucleus, with
a possible further enhancement where $\sigma$ is depressed, suggesting
that the young population is associated with the kinematic
subcomponent.  Along the minor axis the \hbeta\ profile is steeper and
occupies a smaller region, indicating the extent of the young
population is a factor $\sim 0.7$ smaller along the minor axis. If the
young population is confined to a disk, its angle of inclination would
be about $45 \deg$ with respect to the line-of-sight, and curiously
would have approximately the same orientation as the HI ring in the
outer regions \shortcite{schiminovich-etal95}.  The reason why the
possible discontinuity of the \hbeta\ index observed along the major
axis is not apparent along the minor axis may be the caused by a
smaller extent of the young population along this direction, and/or a
larger seeing.

The presence of weak [OIII] emission in our spectra suggests there may
be some hidden \hbeta\ emission too, which could lower the measured
equivalent width, therefore the measured \hbeta\ at the nucleus is
probably a lower limit.

\subsubsection{Mg}\label{sec:mg}

The Mg indices show a gradual increase towards the nucleus, but they
are lowered sharply at the location of the kinematically distinct
core. Along the major axis \mg2 increases from $\sim 0.18$ mag at $r
\approx 16\asec$ to $\sim 0.205$ mag at $r \approx4\asec$, but then
lowers to $\sim 0.153$ at the nucleus.  The \mg1 rises from $\sim
0.06$ mag at $r \approx 16\asec$ to $\sim 0.09$ mag at $r \approx
2\asec$, but then lowers to $\sim 0.037$ at the nucleus.  The same
trend can be seen in the \mgb\ equivalent width, albeit with larger
errorbars. The \mgb\ also has a different shape to the \mg1 and \mg2
at the nucleus, and the depression is shallower and less sharp.  As
discussed earlier in \S\ref{sec:emission}, \mgb\ is more prone to
hidden emission at the nucleus and is probably an upper-limit there.
Similar profiles are observed along the minor axis, but with a noisier
appearance due to lower $S/N$.

The absorption line indices of the bulge could be lowered if the
continuum level is raised by light of very young stars.  The enhanced
\hbeta\ and lowered Mg indices are, in the first instance, consistent
with the picture in which the nuclear light originates from a young
stellar population associated with the kinematically cold
subcomponent, and from an old population associated with the
kinematically hot bulge.

The observed \mg2 indices are more than 0.08 mag lower than that
expected from the $Mg$-$\sigma$ relationship, which predicts $\mg2
\sim 0.294$ mag for $\sigma=200\,\kms$ \cite{davies96iau}. This
difference is too large to be accountable by the possible difference
in zero-points. Together with the high \hbeta\ ($\sim 2 \pm 0.4 \ang$)
of the ``bulge'', this may suggest that some young stars are not
merely confined to the kinematic subcomponent, but is also mixed into
the dynamically hot bulge population.

\subsubsection{Fe}

The \fe{5270} and \fe{5335} profiles are quite different to those of
Mg, with no observable gradient in either index.  This result is
rather surprising given that there are pronounced depressions in the
Mg indices.  If the depressed Mg indices at the nucleus are due to
dilution from a young population with featureless, blue spectra, then
the Fe is expected to be depressed too. Although \fe{5270} could be
enhanced by hidden emission in the continuum bandpass, no significant
depression is seen in \fe{5335}, therefore the flat profiles cannot be
attributed to hidden emission. In the next section we shall quantify
the amount of depression expected in the Fe indices due to light from
a young stellar population.

\section{Can the nuclear line index gradients be explained by light dilution by
O5--A3 stars?}

\begin{figure}
\psfig{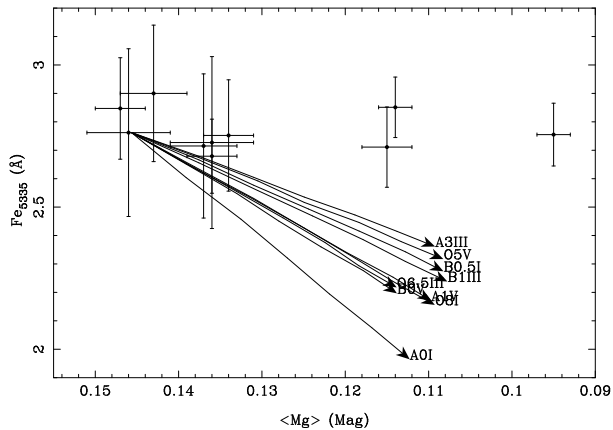}
\caption[]{Effect of light dilution on the \fe{5335} line by adding
spectra of O, B \& A type stars selected from the spectral library of
Jacoby, Hunter \& Christian (1984). The data points along the major
axis of NGC~2865, with $|r| < 4 \asec$, are plotted. The spectrum at
$r= -3.37 \asec$ is extracted and various amount of stellar spectra
between O5 \& A3 are added to it. The arrows denote the trend in which
the \mean{Mg}=(\mg1+\mg2)/2 and \fe{5335} follows as the galaxy
spectrum is diluted more and more by light from the early-type
stars. }
\label{fig:dilution-fe53}
\end{figure}

We want to investigate whether light dilution by young O, B or A type
stars alone are able explain the observed line-strength gradients at
the nucleus of NGC~2865.  For this purpose, the spectrum at $-3.37
\asec$ on the major axis of NGC~2865 is extracted, and varying
proportions of spectra between type O5 and A3 of class I, III and V
taken from a library of flux-calibrated, solar-metallicity spectra are
added to it \cite{JHC84}. \footnote{Although
our spectra are not flux calibrated, this will not change the
conclusions, firstly because the gradients of the lines in
Fig.~\ref{fig:dilution-fe53} are independent of the zero point, and
secondly because the separation between Fe and Mg absorption lines is
small compared with the wavelength range covered.}  The line-indices
are then measured, and in Fig.~\ref{fig:dilution-fe53} the trend which
\fe{5335} and \mean{Mg}=(\mg1+\mg2)/2 follow as a result of light
dilution are plotted over the actual data points at $|r| < 4\asec$
along the major axis.  \footnote{We chose to model \mean{Mg} to boost
signal and to reduce the number of parameters to model. \fe{5335} is
chosen because \fe{5270} {\it could} be affected by hidden
emission. As it turns out, similar results are obtained when
\mean{Fe}=(\fe{5270}+\fe{5335})/2 is modelled instead.}  The
equivalent width of the \fe{5335} line is expected to be lowered by at
least $0.55\ \ang$ for a change of $\sim -0.052$ mag in
\mean{Mg}. This depression is not observed, and it is clear that the
data points do not follow the trend expected from dilution.  Therefore
the primary component of the nuclear light in the core of NGC~2865
does not originate from O5--A3 stars, and the gradient of the spectral
continuum cannot account for the different behavior of the Mg and
Fe. In order to explain the observed line-strength gradients, one must
appeal to a different evolution of individual metallic lines as the
starburst ages---this shall be investigated in the next
section. Results in this section also demonstrate that major star
formation must have ceased for at least o($5 \times 10^8$) yr, the
lifetime of a Main Sequence A3 star.

\section{How do the Mg and Fe absorption lines vary with the Main
Sequence turnoff?} \label{sec:mgfevst} 

\begin{figure}
\centerline{ \hbox{\psfig{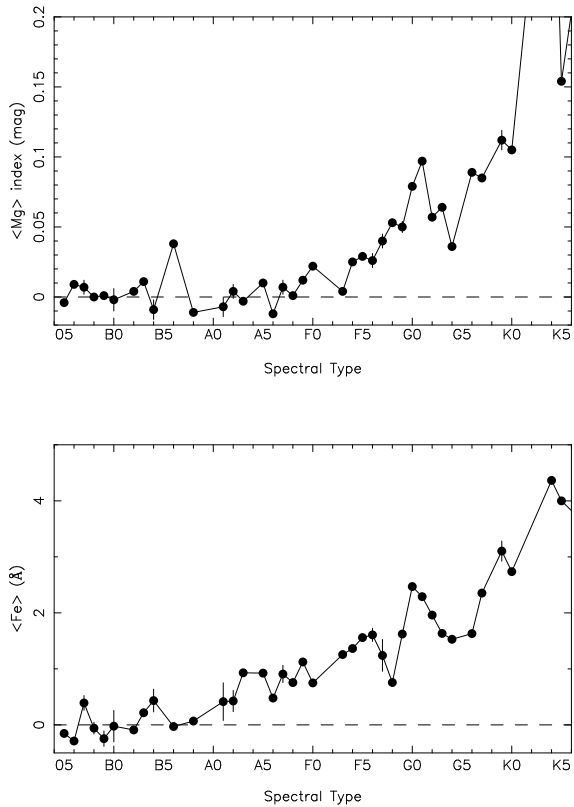}}}
\caption[]{The \mean{Mg} $=(\mg1+\mg2)/2$ and \mean{Fe}
$=(\fe{5270}+\fe{5335})/2$ line-strengths as a function of spectral
type for Main Sequence stars selected from the library of Jacoby \etal
(1984). Where more than one star of the same spectral type is
available, an average and its 1 $\sigma$ errorbar is given.}
\label{fig:mgfe-type}
\end{figure}

During the first 8 Gyr after a starburst, the Main Sequence (MS) light
contribution to the V-band luminosity oscillates with time between 30
and 75\% (assuming Salpeter IMF; Fig.~8 \citeNP{CB91}). Furthermore, during this time the spectral energy distribution
of the MS will approximate that of the stars at the turn-off, because
their brightness outweighs the number of stars at later types.  How do
the Mg and Fe indices behave as a function of spectral type?  Line
indices are measured for a sample of standard solar-metallicity stars
with spectral class V \shortcite{JHC84} and plotted
against the spectral type in Fig.~\ref{fig:mgfe-type}.  Spectra of MS
stars earlier than A3 do not have a significant signal in \mean{Mg}
and \mean{Fe}. Between approximately A3 and F4, the \mean{Fe} is
significant, whilst \mean{Mg} is still negligible.  Only after about
F4 are both \mean{Mg} and \mean{Fe} significant.

The ages of an instantaneous burst with MS turn-off at A3 and F4 are
estimated in the following way: Taking (B-V) of solar metallicity A3V
and F4V stars to be 0.09. and 0.39 respectively \cite{MB81}, their corresponding effective temperatures are $8760$ K and
$6700$ K respectively (\citeNP{MB81},
\shortciteNP{alonso-etal96}).  By linearly interpolating the
theoretical isochrones of \shortciteN{bertelli-etal94}, we arrive at
age estimates of 0.5 Gyr and 1.7 Gyr for MS turn-offs at A3 and F4
respectively. This result is consistent with Bica \& Alloin's (1987)
age determination using star cluster templates.

In summary, about 0.5--1.7 Gyr after an instantaneous burst, the Fe
absorption lines in the burst population may be stronger than that of
the Mg.  This is due to a temperature effect in the atmospheres of
stars at the MS turnoff.  When the spectrum of this population is
added to that of the old population, the total \mean{Mg} will be
lowered more than the \mean{Fe}, and the inferred [Mg/Fe] abundance
ratio appears to be modified.  This difference may be enhanced by
increasing the light contribution from the burst population (i.e. by
increasing its strength), by increasing the overall metallicity of the
burst, or by increasing its Fe relative to Mg.  Indeed, population
synthesis in the next section shows that an enhancement in metallicity
or Fe abundance is required to explain the nuclear line indices of
NGC~2865.  Note that this effect only applies if the age spread of the
newly formed stars is smaller than the MS age difference between A3 \&
F4 (i.e. $<< 1.2$ Gyr).  We also note that although the difference in
the Mg and Fe indices can be attributed to the temperature effects of
the stars at MS turnoff, if the metallicity is high enough, after
about 1 Gyr the Mg and Fe indices in the burst population are
comparable to those of the old population, thus kinematics
measurements can be performed (\S~\ref{sec:kinematics}). In the next
section, the observed indices are compared with predictions by
evolutionary stellar population synthesis.

\section{Constraints on the recent star formation history by stellar 
population synthesis}
\label{sec:bcmodelling}

\begin{figure}
\psfig{figure=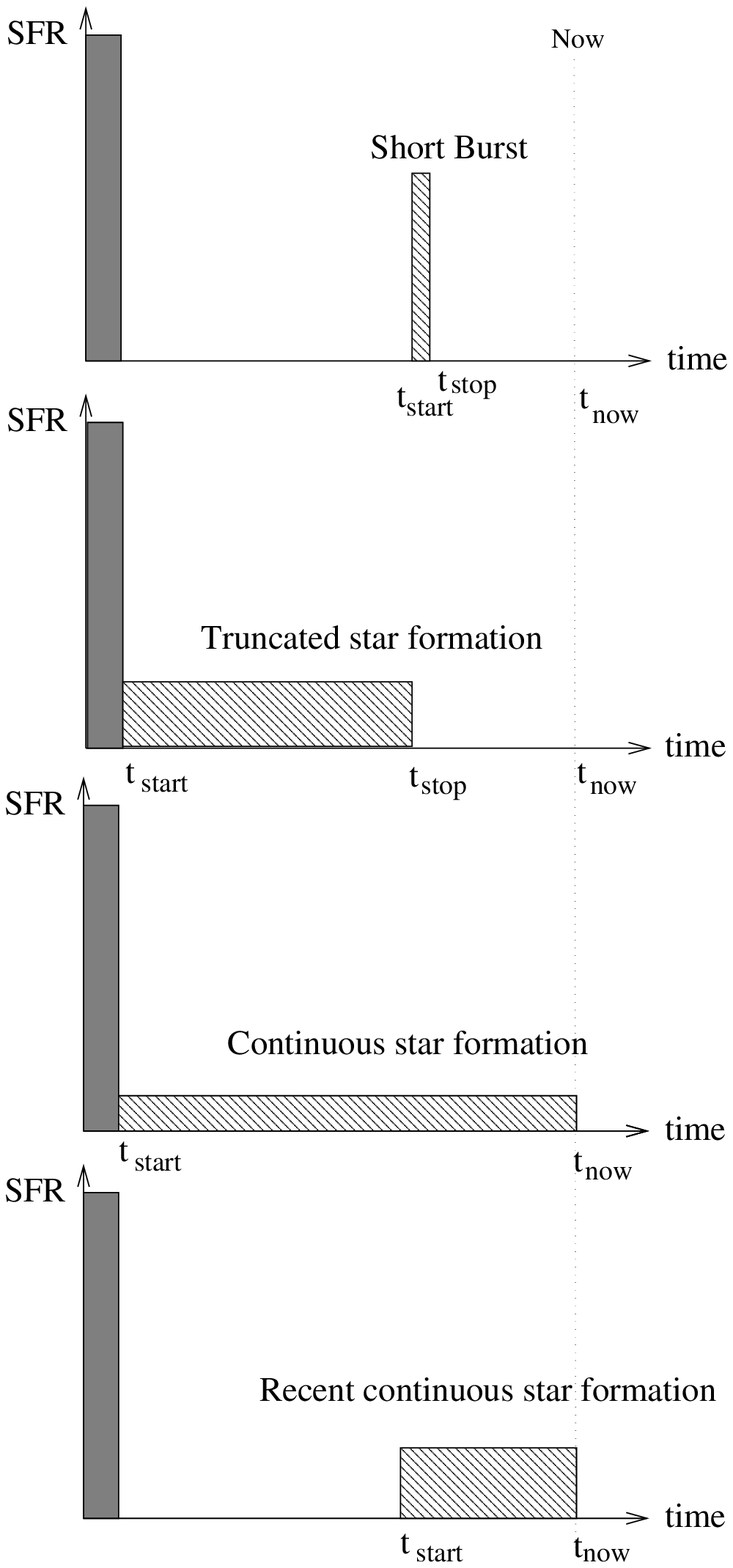,width=7cm}
\caption[]{The 4 different star formation scenarios considered. The
abscissa is the age of the galaxy and the ordinate is the star
formation rate. The star formation rate of the galaxy bulge is shaded
in grey and that of the nuclear component is shaded in black. }
\label{fig:models}
\end{figure}

\begin{table*}
\centering
\caption{A summary of the parameters adopted in the star formation
models. The first 6 columns are, respectively: model name, the start
and stop time of the secondary episode of star formation in Gyr, the
metallicity of the bulge and the nuclear subcomponent, and the [Fe]
overabundance (in dex) of the nuclear component relative to solar. The
remaining 2 columns summarize the model parameters which satisfy the
observational constraints. They are the time elapsed since cessation
of the secondary episode of star formation $\Delta t=t_{now} -
t_{stop}$, and the ``burst fraction'' $f_b$ along the line-of sight at
the nucleus.}
\label{tab:runs}
\vspace{0.5cm}
\begin{tabular}{llllllll}
\hline
Model & $t_{start}$ & $t_{stop}$ & $Z_{bulge}(\zsun)$ & 
$Z_{nuc}(\zsun)$ & $\Delta$ [Fe] (dex)  & $\Delta t$ (Gyr)& $f_b$ \\
\\
SB1  &7 &7.1 &1  &1  & 0    &-    &-           \\
SB2  &18&18.1&0.4&0.4& 0    &-    &-           \\
SB3  &7 &7.1 &1  &2.5& 0    & 0.7 & 0.33       \\        
     &  &    &   &   &      & 0.8 & 0.50       \\        
     &  &    &   &   &      & 0.9 & 0.67--0.91 \\        
     &  &    &   &   &      & 1.0 & 0.91       \\        
SB4  &18&18.1&0.4&1  & 0    & 1.0 & 0.20       \\
     &  &    &   &   &      & 1.2 & 0.33       \\       
     &  &    &   &   &      & 1.3 & 0.33       \\
     &  &    &   &   &      & 1.4 & 0.50       \\ 
     &  &    &   &   &      & 1.5 & 0.50       \\
     &  &    &   &   &      & 1.6 & 0.67--0.91 \\
     &  &    &   &   &      & 1.7 & 0.91       \\ 
SB5  &18&18.1&0.4&2.5& 0    & 0.4 & 0.13       \\ 
     &  &    &   &   &      & 0.5 & 0.20--0.33 \\       
     &  &    &   &   &      & 0.6 & 0.20--0.91 \\       
     &  &    &   &   &      & 0.7 & 0.33--0.91 \\       
     &  &    &   &   &      & 0.8 & 0.50--0.91 \\       
     &  &    &   &   &      & 0.9 & 0.91       \\       
SB6  &7 &7.1 &1  &1  & 0.08  & -   & -          \\
SB7  &7 &7.2 &1  &2.5& 0    & 0.6 & 0.20       \\  
     &  &    &   &   &	    & 0.7 & 0.33--0.50 \\
     &  &    &   &   &	    & 0.8 & 0.50--0.91 \\      
     &  &    &   &   &	    & 0.9 & 0.67--0.91 \\
     &  &    &   &   &	    & 1.0 & 0.91       \\      
SB8  &7 &7.4 &1  &2.5& 0    & 0.5 & 0.20       \\
     &  &    &   &   &	    & 0.6 & 0.33       \\
     &  &    &   &   &	    & 0.7 & 0.50--0.67 \\
     &  &    &   &   &	    & 0.8 & 0.67--0.91 \\
     &  &    &   &   &	    & 0.9 & 0.91       \\
TSF1 &1 &7   &1  &1  & 0    &-    &-           \\
TSF2 &1 &7   &1  &1  & 0.04  &-    &-           \\
TSF3 &1 &7   &1  &1  & 0.08  & 0.1 & 0.50       \\
     &  &    &   &   &	    & 0.2 & 0.67--0.80 \\ 
     &  &    &   &   &	    & 0.3 & 0.80--0.99 \\
     &  &    &   &   &	    & 0.4 & 0.99       \\    
CSF1 &1 &-   &1  &1  & 0    &-    &-           \\
CSF2 &1 &-   &0.4  &0.4  & 0    &-    &-           \\
RCSF1&7 &-   &1  &1  & 0    &-    &-           \\
RCSF2&7 &-   &1  &2.5& 0    &-    &-           \\
\hline
\end{tabular}
\end{table*}

We shall constrain the recent star formation history of NGC~2865 using
population synthesis, by considering a ``bulge+burst'' model in
which nuclear light originates from an old bulge population plus a
younger population associated with the kinematically distinct
component.\footnote{For convenience, throughout this work we shall
refer to the separate, younger population as the ``burst'', even
though it may have prolonged star formation.}  Furthermore, the star
formation histories are simplified by considering 4 basic shapes,
illustrated in Fig.~\ref{fig:models}.

Due to degeneracy, the details of star formation cannot be
distinguished after a few Gyr. So for the purpose of this work, the
bulge stars are formed between 0 \& 1 Gyr after the Big Bang with a
constant star formation rate $SFR$.  Initially, the bulge \hbeta\ is
strong due to the dominance of young, bright early-type stars, whilst
the metallic line-strengths are weak. As the bulge ages, its \hbeta\
falls whilst its metal line-indices rise and approach some asymptotic
values similar to those observed at present.  The introduction of a
secondary episode of star formation (the ``burst'') immediately lowers
the metallic line indices of the combined nuclear light and raises its
\hbeta, due the presence of early-type stars. Whilst star formation is
ongoing, \hbeta\ remains high
\footnote{Ignoring the presence of emission lines when the star
formation rate is high}, whilst the metallic line indices remain
lowered.  After star formation stops, the line-indices will again
approach some asymptotic values. The \hbeta\ of the combined nuclear
light approaches that of the bulge from above. The asymptotic values
of the metallic line indices depend on the metallicity $Z_{nuc}$ of
the new stars. In the absence of self-enrichment, they would be higher
than the bulge values if $Z_{nuc}$ is higher than the bulge
metallicity $Z_{bulge}$, and vice versa.

The {\it absolute} values of our line-indices cannot be modelled due
to uncertainty in their zero-points. Therefore we model the line-index
{\it differences} between the combined nuclear light
(i.e. bulge+burst) and that of the bulge. Furthermore we concentrate
on modelling $\Delta \hbeta$, $\Delta \mean{Mg}$ \& $\Delta
\fe{5335}$, because \mgb\ and \fe{5270} could be affected by weak
emission, as discussed in \S\ref{sec:emission}. We adopt
$\Delta\mean{Mg}$ between the nucleus and the bulge to be
$\Delta\mean{Mg} = -0.053 \pm 0.012$ mag, and $\Delta\fe{5335} = 0.00
\pm 0.35 \ang$.  Because of possible hidden \hbeta\ emission,
indicated by weak [OIII] emission, $\Delta\hbeta$ is taken to be
$\Delta\hbeta \age 1.4 \ang$. Star-formation rates which satisfy these
observational constraints are sought.

Star formation rates ($SFR$) of the 4 basic shapes illustrated in
Fig.\ref{fig:models} are investigated, with the adopted parameters
summarized in Table~\ref{tab:runs}.  In models SB1--SB9 we investigate
whether the nuclear subcomponent is formed in a recent starburst which
may be triggered by, for example, a spiral-elliptical merger
\cite{HW92}. These models can also apply to
spiral-spiral mergers as long as the stars are well segregated into
two distinct populations, that of the older and dynamically hot
``bulge'' consisting of stars formed before the collision, and the
young compact nuclear ``burst'' population formed from gas during the
collision. We note in this case the bulge stars on average would
appear to be younger than the Universe. Next, in models TSF1--TSF3, we
investigate whether the nuclear stars were accreted from a
star-forming galaxy in an event which also truncated the star
formation in this population; the feasibility of this model is
discussed in \S~\ref{sec:discussion}.  Next, in model CSF1, we
investigate whether the nuclear subcomponent is primordial and formed
by constant low-level star formation over a Hubble time up to the
present. This model provides a simple approximation to small embedded
disks in a ``disky'' elliptical, proposed to be a continuation of the
morphological sequence of spirals into ellipticals
\cite{scorza-bender90}.  Finally in models RCSF1 \& RCSF2, we
investigate whether the nuclear component was formed recently by
continuous and ongoing infall of cold gas that could be returned tidal
material after a spiral-spiral merger (Hibbard \& Mihos 1995).  The
range of model parameters which satisfy the observational constraints
are summarized in the last 2 columns of Table~\ref{tab:runs}.

The Galaxy Isochrone Synthesis Spectral Evolution Library (GISSEL96)
of \citeN{BC97} is used as a diagnostic tool. It
follows the evolution of a stellar population along the evolutionary
tracks of the Padova School, up to an age of 20 Gyr, with non-solar
metallicities modelled by Kurucz's (1995 version) model atmospheres.
The library we use is restricted to metallicities of 0.4, 1 \& $2.5\
\zsun$ ([Fe/H]=-0.330, 0.093 \& 0.560 respectively) and a Scalo
IMF. Although in reality the IMF depends on the details of
star-formation, there is still little information in this subject
area, therefore we use a Scalo IMF and discuss the likely effect of
changing it later in \S~\ref{sec:discussion}.

To match the bulge Mg and Fe line-indices to the observed levels, a
bulge age of about 8 Gyr is needed for $Z_{bulge}= 1\ \zsun$, and
about 19 Gyr for $Z_{bulge}=0.4 \zsun$.  Because of uncertainties in
the age of the Universe, bulges with metallicity of $1.0 \zsun$ \&
$0.4 \zsun$ are modelled.

\subsection{Recent, short burst of star formation (SB1-SB8)}

\begin{figure*}
\psfig{figure=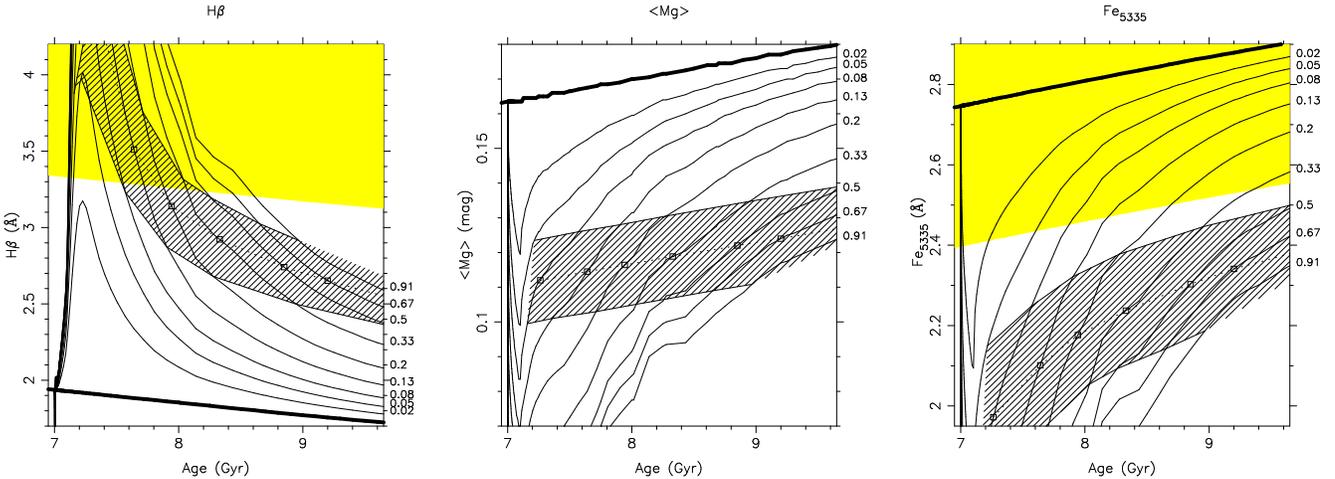,angle=270,width=17.5cm}
\caption[]{The \hbeta, \mean{Mg} and \fe{5335} for model SB1 where the
nuclear subcomponent is formed in a 0.1 Gyr starburst with the same
(solar) metallicity as the bulge. The ordinate is the line index and
the abscissa is the galaxy age. The bulge is formed between 0 \& 1 Gyr
and its line indices are plotted with thick solid lines.  The nuclear
component begins to form at 7 Gyr.  Labelled along the right edges are
the burst fraction $f_b$ of the starburst.  The observational
constraints are shaded.  In the central panel, the hatched region
enclosing a dashed curve with square symbols indicates a nuclear
\mean{Mg} lowered by $0.053 \pm 0.012$ mag relative to the bulge. This
same region (hatched) is mapped into the left and right panels for
comparison purpose.  In addition, the region corresponding to a
nuclear \hbeta\ enhanced by $\ge 1.4 \ang$ relative to the bulge is
shaded in the left panel, whilst in the right panel, the region
corresponding to $|\Delta \fe{5335}| \le 0.35 \ang$ is shaded. In
order to have consistent solutions, the shaded and hatched regions
must overlap in the left and right panels {\it at the same age and
$f_b$}.}
\label{fig:sb-samez}
\end{figure*}

\begin{figure*}
\psfig{figure=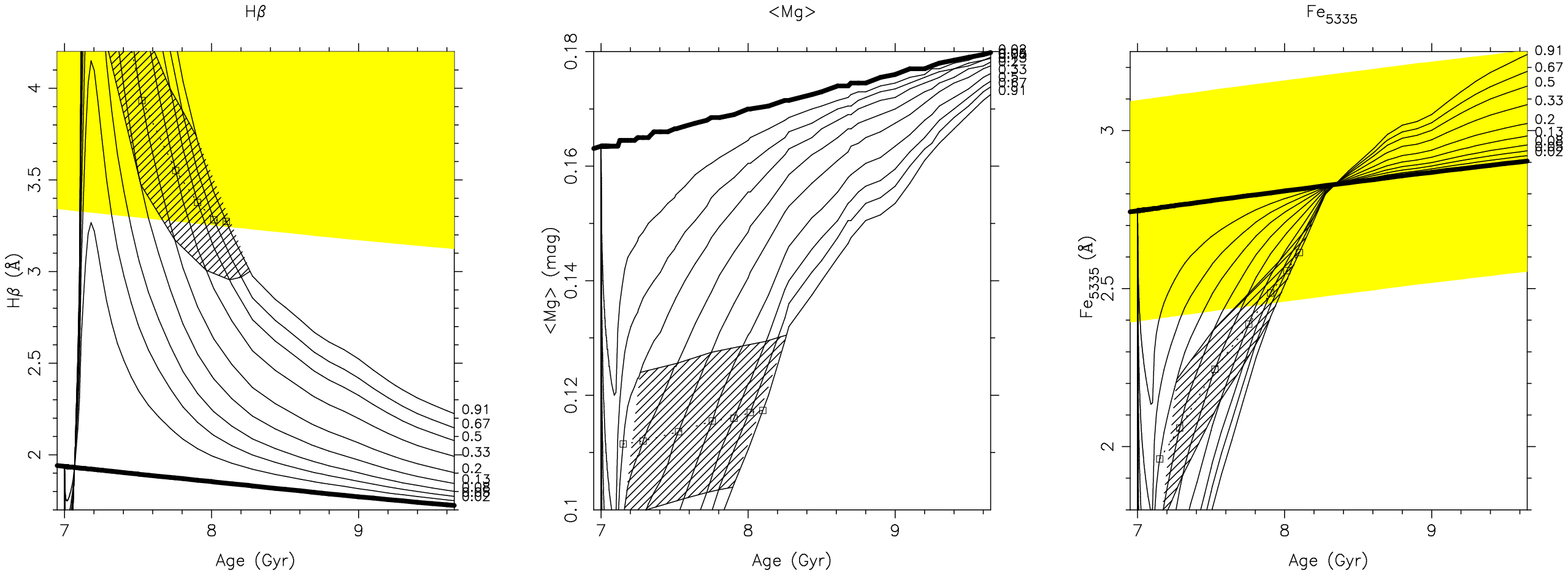,angle=270,width=17.5cm}
\psfig{figure=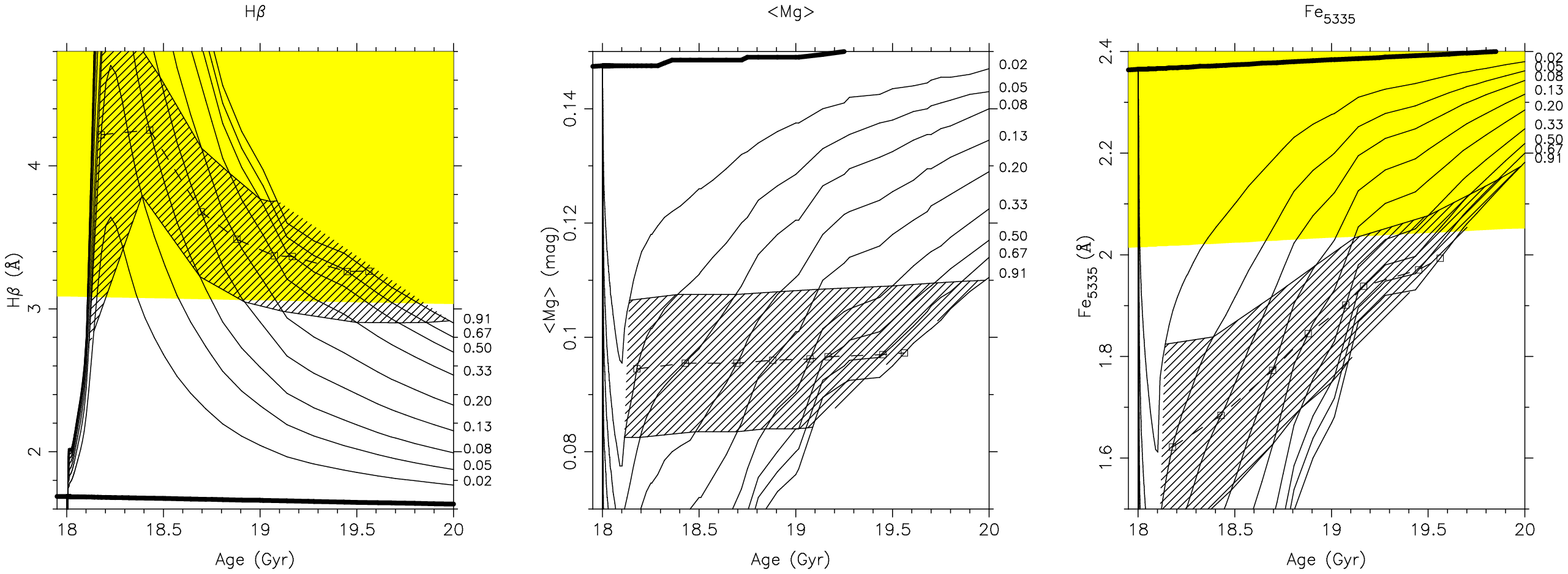,angle=270,width=17.5cm}
\psfig{figure=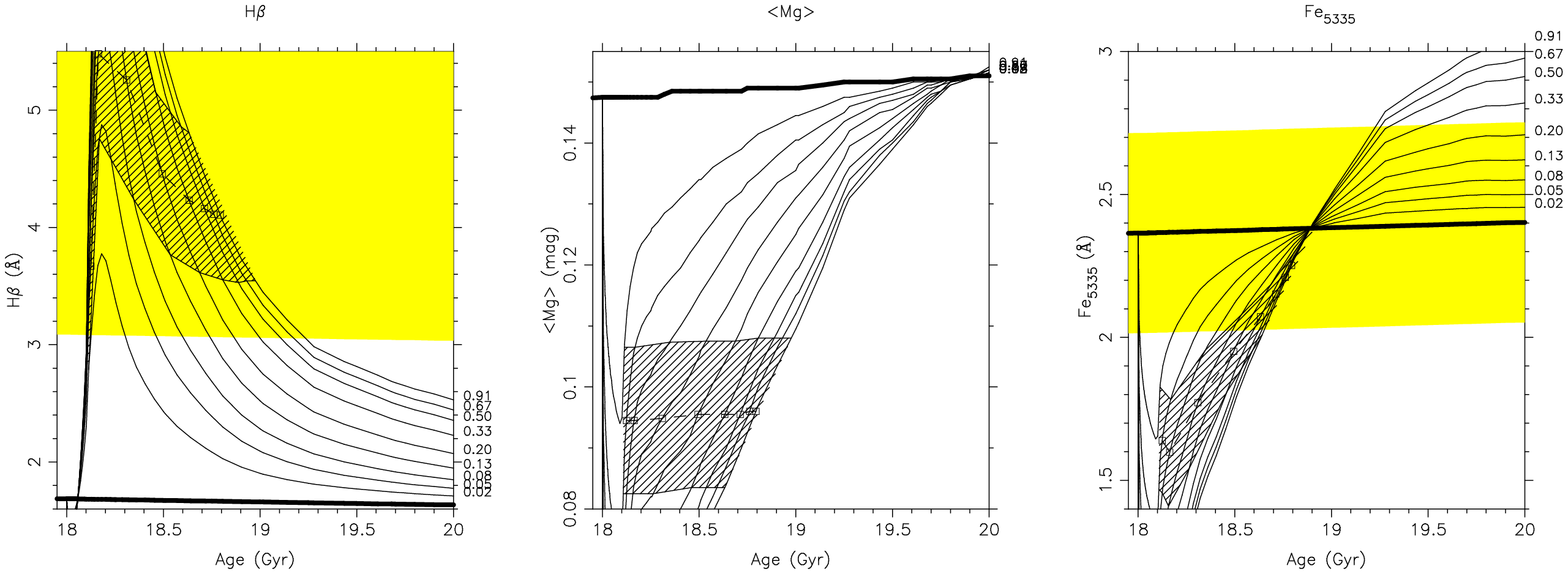,angle=270,width=17.5cm}
\caption[]{The \hbeta, \mean{Mg} and \fe{5335} for models SB3--SB5
where the nuclear subcomponent is formed in a 0.1 Gyr burst with
metallicity higher than that of the bulge.  The top row corresponds to
$Z_{bulge} = 1 \zsun$ and $Z_{nuc}=2.5\ \zsun$ (model SB3), the middle
row corresponds to $Z_{bulge} = 0.4\ \zsun$ and $Z_{nuc}=1\ \zsun$
(model SB4), whilst the bottom row corresponds to $Z_{bulge} = 0.4\
\zsun$ and $Z_{nuc}=2.5\ \zsun$ (model SB5). Labelled along the right
edges are the burst fraction $f_b$ of the starburst. The observational
constraints are shaded; see Fig~\ref{fig:sb-samez} for explanation.}
\label{fig:sb-enhancez}
\end{figure*}

Could the nuclear subcomponent be formed in a recent starburst? We
form the nuclear subcomponent at $t_{{\rm start}}$ in a starburst
which lasted 0.1 Gyr.  The metallicity of the burst population can be
the same or higher than that of the bulge. A subcomponent with
$Z_{nuc} < Z_{bulge}$ is not modelled because large $\Delta \mean{Mg}$
\& $\Delta \fe{5335}$ are achievable with low SFR and gives small
$\Delta \hbeta$.  The starburst strength is parameterized by burst
fraction $f_b$, defined as the ratio of burst mass to total mass of
the system along the line-of-sight.

\subsubsection{Starbursts with the same metallicity as the bulge
population (SB1, SB2)} \label{sec:samez}

In Fig.~{\ref{fig:sb-samez}} the line-indices for model SB1 where
$Z_{nuc}=Z_{bulge}=1\ \zsun$ are presented. Shaded are also the
observational constraints.  The observed $\Delta \hbeta$ \& $\Delta
\mean{Mg}$ are reachable within 1.0 Gyr after the starburst, but the
predicted $|\Delta \fe{5335}| \age 0.5 \ang$ is ruled out by
observations at 92\% confidence level.  We find model SB2, where the
metallicity is $0.4\ \zsun$, is also ruled out by observations, and
the results are not plotted.

\subsubsection{Starbursts with metallicity higher than the bulge
population (SB3--SB5)}

The above models are repeated with $Z_{nuc} > Z_{bulge}$. The
line-indices are presented in Fig.~\ref{fig:sb-enhancez}, where the
top row of diagrams corresponds to $Z_{bulge} = 1\ \zsun$ \& $Z_{nuc}
= 2.5\ \zsun$, the middle row corresponds to $Z_{bulge} =0.4\ \zsun$
\& $Z_{nuc}= 1\ \zsun$, and the bottom row corresponds to $Z_{bulge}
=0.4\ \zsun$ \& $Z_{nuc}= 2.5\ \zsun$.  The best-fit parameters are
summarised in Table~\ref{tab:runs}.

Since $Z_{nuc} > Z_{bulge}$, the metal line-indices after the
starburst rise above the asymptotic values of the bulge.  The
\fe{5335} rises faster than \mean{Mg}, and crosses over the bulge
value earlier, at 0.9--2.5 Gyr after the starburst, confirming our
findings in \S~\ref{sec:mgfevst}. If $Z_{bulge}=1 \zsun$ and $Z_{nuc}=
2.5\ \zsun$, solutions exist some 0.7--1.0 Gyr after the starburst,
for $0.33 \le f_b \le 0.91$.  If $\Delta t = t_{now}-t_{stop} = 0.7$
Gyr, a $f_b= 0.33$ is required.  If $\Delta t = 0.9$ Gyr, $0.67\le f_b
\le 0.91$ is required. If $\Delta t = 1.0$ Gyr, a high $f_b=0.91$ is
required.

If $Z_{bulge}= 0.4\ \zsun$, solutions exist for both $Z_{burst}= 1\
\zsun$ \& $2.5\ \zsun$, with the latter satisfying the constraints
better.  If $Z_{burst}= 1\ \zsun$, the observational constraints can
be satisfied if a burst with $0.2 \le f_b \le 0.91$ stopped between
1.0 \& 1.7 Gyr ago (see Table~\ref{tab:runs}). If $Z_{burst}$ is
raised to $2.5\ \zsun$, a more recent burst 0.4--0.9 Gyr ago with
$0.13 \le f_b \le 0.91$ is needed. If $\Delta t = 0.4$ Gyr, $f_b=
0.13$ is required. If $\Delta t = 0.6$ Gyr, $0.20 \le f_b \le 0.91$ is
required, whilst if $\Delta t = 0.8$ Gyr, $0.50 \le f_b \le 0.91$ is
required. Finally, if $\Delta t = 0.9$ Gyr, a rather high $f_b$ of
0.91 is required; interestingly at this time the nuclear \fe{5335} is
$\sim 0.1 \ang$ {\it higher} than that of the bulge.

In summary, solutions exist if the nuclear subcomponent is made by a
starburst with metallicity $\sim 2.5$--6 times that of the bulge
population and nuclear line-of-sight $f_b$ ranging from 0.13 to 0.91.
Low $f_b$'s are needed for recent bursts and {\it vice versa}. Extreme
$f_b$'s satisfy the constraints marginally whilst $f_b$ between about
0.33 \& 0.91 give better fits.  The burst mass may be lowered if its
metallicity is raised, as discussed in \S~\ref{sec:mgfevst}. Age
estimates range from 0.4 Gyr to 1.7 Gyr, depending on $f_b$,
$Z_{bulge}$ \& $Z_{nuc}$.  Tighter constraints would require better
data, although the parameters would somewhat be limited by degeneracy.

\subsubsection{Starbursts with enhanced [Fe] abundance (SB6)} 

\begin{figure*} 
\psfig{figure=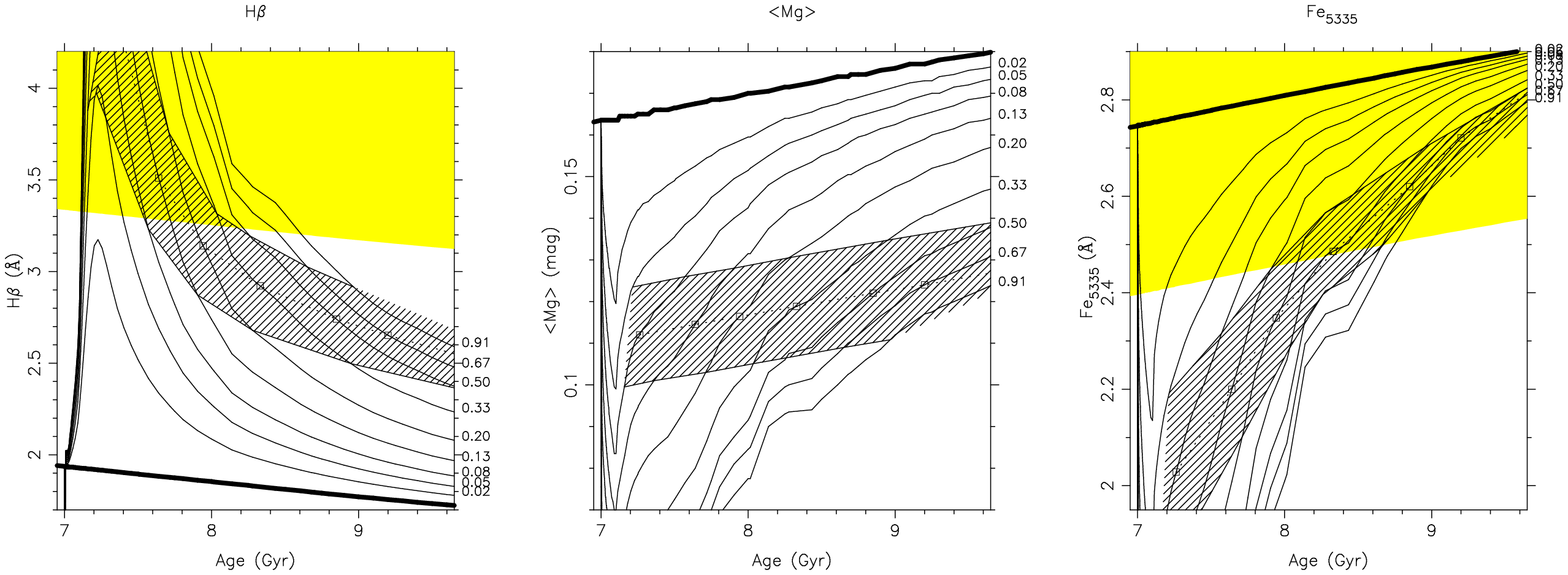,angle=270,width=17.5cm}
\caption[]{The line indices for model SB7 where the nuclear component
is formed in a 0.1 Gyr starburst at 7 Gyr, with the same (solar)
metallicity as the bulge.  However, the bulge Fe abuncance is 0.08
dex over the solar value (model SB7). Labelled along the right edges
are the burst fraction $f_b$ of the starburst. The observational
constraints are shaded; please see Fig~\ref{fig:sb-samez} for
explanation. }
\label{fig:efe} 
\end{figure*}

If the kinematic subcomponent is formed from gas accreted from a
spiral into a giant elliptical, it may have a [Mg/Fe] abundance ratio
lower than that of the underlying elliptical, because giant
ellipticals are thought to have an over-abundance in light-elements
(or under-abundance in heavy elements).  This is further discussed in
\S~\ref{sec:tsfefe} \& \S~\ref{sec:discussion}.

There are 2 ways of approach. The first is to keep the Fe abundances
in the 2 galaxies the same, and raise the Mg in the elliptical.  This
will not work as too much depression in \mean{Mg} and \fe{5335}, and
too little \hbeta\ enhancement is predicted. The second way, which is
adopted, is to keep the Mg abundances the same, but raise the Fe in
the burst population. In this case $Z_{nuc}$ is actually slightly
higher than $Z_{bulge}$. Because stellar populations in GISSEL96 have
solar [Mg/Fe] abundances, we artificially enhance the burst \fe{5335}
by decreasing the flux in model passbands, whilst ignoring the
physical processes involved. Fig.~\ref{fig:efe} show the results for
model SB6 where $Z_{nuc} = Z_{bulge} = 1 \zsun$, with the burst Fe
enhanced by 0.08 dex over solar (model SB6).  We find this model is
marginally ruled out. Although it may be saved by increasing the Fe
abundance by more than 0.08 dex, it is probably difficult to increase
by much, as fig.~1 of \cite{davies96iau} shows that the typical
abundance difference is $\sim 0.1$ dex.

\subsubsection{Constraints on the burst duration (SB7, SB8)}

\begin{figure*}
\psfig{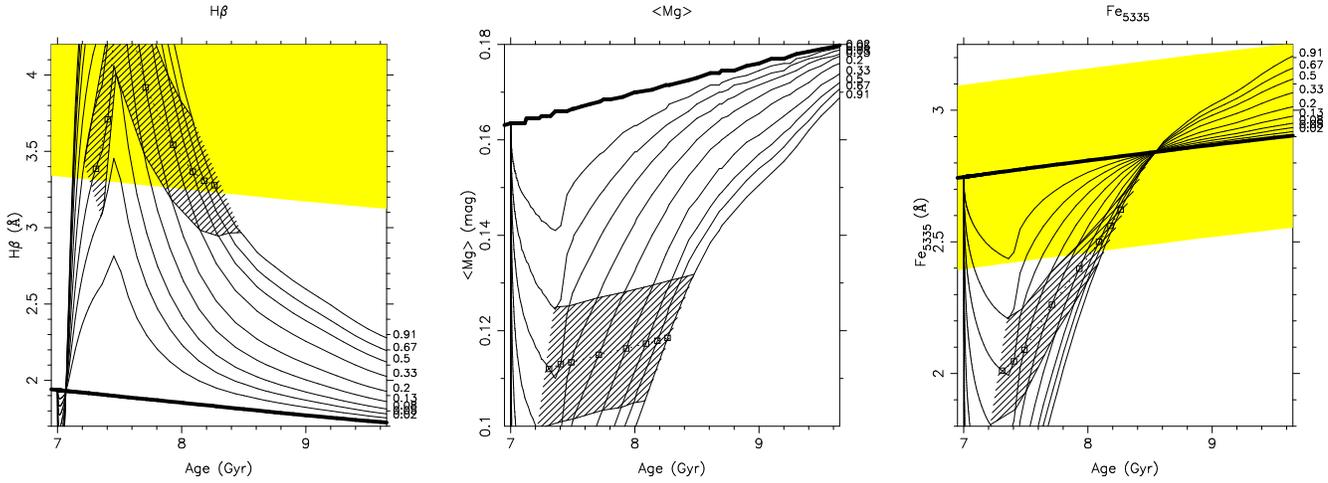}
\caption[]{The line indices for model SB8 where the nuclear component
is formed in a starburst which started at 7 Gyr, and lasted 0.4
Gyr. The burst has metallicity of 2.5 \zsun whilst the bulge has solar
metallicity. }
\label{fig:burstwidth}
\end{figure*}

The burst duration is constrained by repeating model SB3 with burst
durations of 0.2 and 0.4 Gyr, respectively, in models SB7 \& SB8. The
line-indices for the latter are plotted in
Fig.~\ref{fig:burstwidth}. The observational constraints can be
marginally met 0.5 Gyr after the burst with $f_b= 0.20$, after 0.6 Gyr
with $f_b = 0.33$, or after 0.9 Gyr with $f_b=0.91$.  The burst
duration can be constrained weakly, and no solution exists for
durations longer than $\sim 0.4$ Gyr. Due to the short lifetimes of
A-type stars there is less \hbeta\ for a given $f_b$. As a consequence
of increasing the burst duration, the observation has to be made
earlier, and furthermore the range of permitted $f_b$'s becomes more
restricted.

We conclude that the burst duration can only be weakly constrained.
Shorter bursts are favored because the range of $f_b$ is less
restricted. We also note that because the orbital period of the
secondary galaxy is an order of magnitude lower than the maximum burst
duration, it will probably be impossible to distinguish a succession
of bursts from a single burst.

\subsection{Truncated star formation (TSF1-TSF3)}

\begin{figure*}
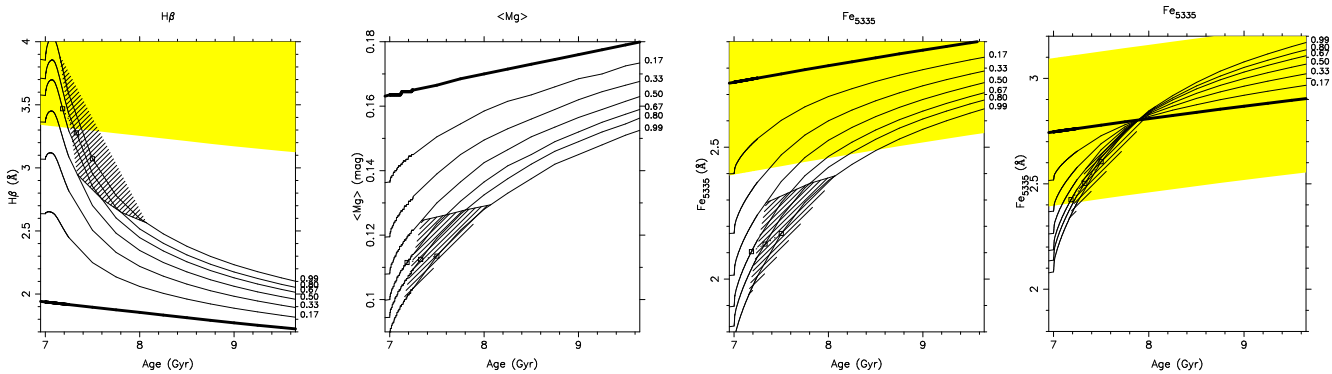

\hbox{\psfig{figure=fig12lef.ps,angle=270,width=13.3cm}
\psfig{figure=fig12ri.ps,angle=270,width=4.2cm}}
\caption[]{{\it Left 3 panels:} The \hbeta, \mean{Mg} and \fe{5335}
for model TSF1 where the nuclear subcomponent is formed by constant
star formation, truncated at $7$ Gyr. The metallicity of the nuclear
subcomponent is the same as that of the bulge, which is solar.
Labelled along the right edges are the burst fraction $f_b$ along the
line-of-sight, defined as the mass of the nuclear component divided by
the total mass. {\it Right panel:} The \fe{5335} for model TSF3 in
which the accreted stars have a Fe abundance of 0.08 dex over
solar. The \hbeta\ and \mean{Mg} in this model remain the same as
those in model TSF1.  }
\label{fig:tsf-samez}
\end{figure*}

Next we investigate whether the nuclear subcomponent is formed by
continuous star formation that has been truncated recently.  The
feasibility of this model is further discussed in
\S~\ref{sec:discussion}.  The stars in the nuclear subcomponent are
formed between 1 \& 7 Gyr with the same, solar metallicity as the
bulge.  The indices are plotted in Fig.~\ref{fig:tsf-samez} for model
TSF1. The strength of star-formation along the line of sight is
parameterized by $f_b$, redefined as the mass of the nuclear
subcomponent divided by the total mass. The observational constraints
cannot be met. Even with $f_b = 0.99$, the desired $\Delta \hbeta$ is
only attainable some 0.2 Gyr after $t_{stop}$, whilst the $\Delta
\fe{5335}$ of $-0.75 \ang$ is ruled out at 98\% confidence level. A
high mass is required because only the recently formed stars are
responsible for lowering the nuclear metallic line indices. As a
consequence too many stars are made during prolonged star
formation. This problem will be worse if the bulge has sub-solar
metallicity so that the galaxy is older.

This model cannot be saved by lowering or increasing the metallicity
of the accreted stars either. If $Z_{nuc} < Z_{bulge}$, the desired
$\Delta \mean{Mg}$ can be reached with a smaller $f_b$, but there are
not enough young stars to give the desired $\Delta \hbeta$. In
addition, $\Delta \fe{5335}$ will be too large.  If $Z_{nuc} >
Z_{bulge}$, $\Delta \fe{5335}$ will be smaller, but an unrealistically
high $f_b$ is required to give the desired $\Delta \mean{Mg}$, because
$\Delta \mean{Mg}$ is pushed up too by the increased metallicity. This
is confirmed when model TSF1 is repeated with $Z_{nuc}= 2.5 \zsun$,
which shows that no solution exists for this model.

\subsubsection{Enhanced Fe abundance in the accreted stars (TSF2, TSF3)}
\label{sec:tsfefe}

Giant ellipticals tend to have Mg indices enhanced relative to Fe (or
Fe is depleted relative to Mg), whilst spiral galaxies tend to have
solar [Mg/Fe] abundance ratios.  If the kinematic subcomponent consist
of stars accreted from a spiral into a giant elliptical, it is likely
to have a [Mg/Fe] ratio lower than that of underlying elliptical. We
find a difference in [Mg/Fe] abundance ratio alone can explain the
observations.  This is demonstrated when the above model is repeated
with the Fe abundance of the nuclear subcomponent artificially
enhanced by 0.04 \& 0.08 dex respectively (models TSF2 \& TSF3),
whilst keeping the metallicity in both components solar. Model TSF2 is
marginally ruled out by observations. The \fe{5335} indices for model
TSF3 are plotted in the right panel of Fig.~\ref{fig:tsf-samez},
showing that an Fe enhancement of 0.08 dex can meet the observational
constraints, although rather high $f_b$'s are required. Solutions
exist at $\Delta t = 0.2$ Gyr for $0.67 < f_b < 0.80$, or at $0.4$ Gyr
for $f_b=0.99$. Therefore a starburst is unnecessary if the nuclear
population has a Fe abundance enhanced by 0.08 dex, and are accreted
by a process which truncated its star formation. A rather high $f_b
\age 0.50$ is required because only the stars made recently are
responsible for the enhanced \hbeta, thus many stars are made over a
Hubble time.

\subsection{Continuous star formation up to the present (CSF1, CSF2)}

\begin{figure*}
\psfig{figure=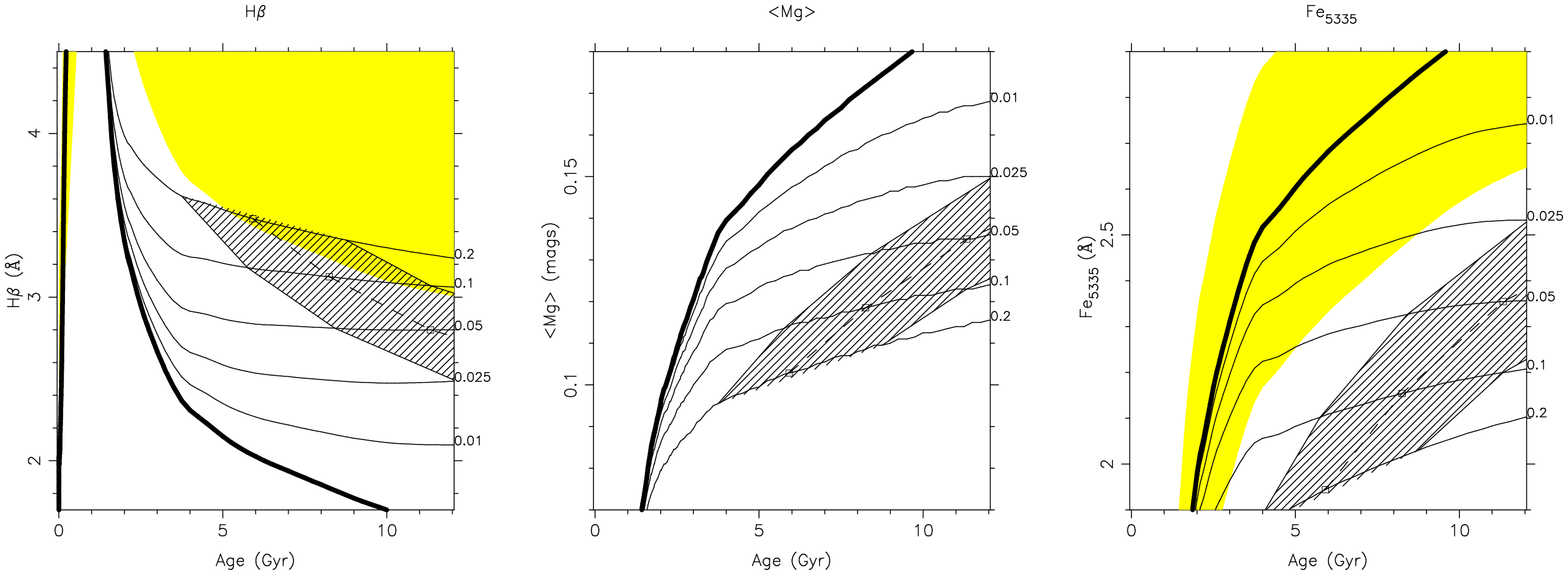,angle=270,width=17.5cm}
\caption[]{The \hbeta, \mean{Mg} and \fe{5335} for model CSF1 where
the nuclear subcomponent is formed by continuous star formation up to
the present. Both the nuclear subcomponent and the bulge have solar
metallicity.  The bulge (which has a mass of $1 \msun$) is formed
between $0$ \& $1$~Gyr. The nuclear subcomponent is formed by constant
star formation from $1$ Gyr onwards. Labelled along the right edges in
units of $\msun/{\rm Gyr}$ are the normalised star formation rates.}
\label{fig:csf000-004}
\end{figure*}

\begin{figure*}
\psfig{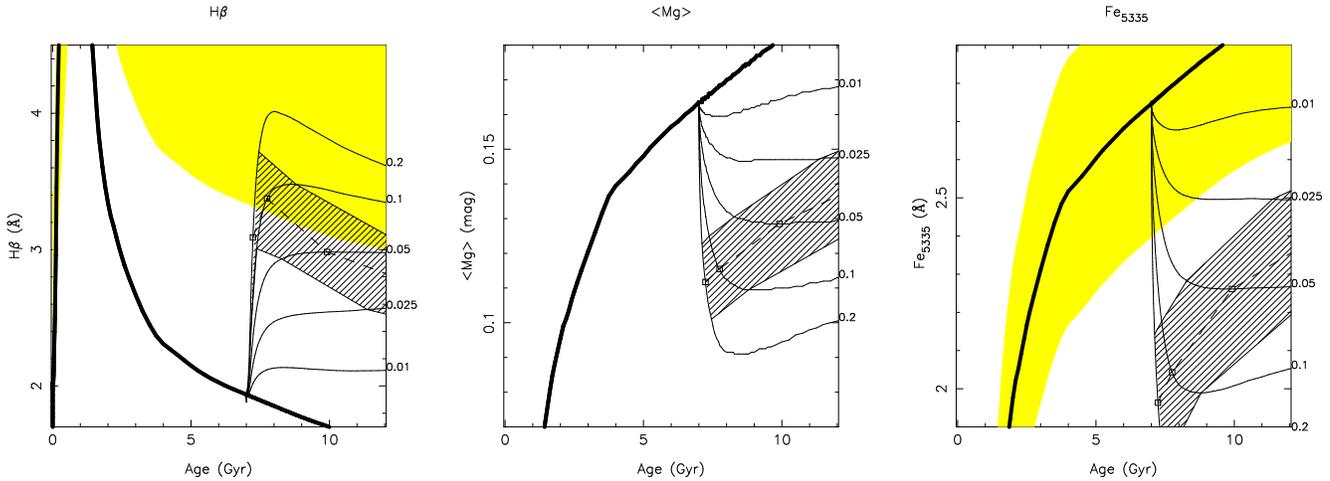}

\caption[]{The \hbeta, \mean{Mg} and \fe{5335} for model RCSF1 where
the nuclear subcomponent is formed by star formation that begins
recently and continues up to the present. The bulge, which has a mass
of $1 \msun$, is formed between $0$ \& $1$~Gyr, whereas the nuclear
subcomponent is formed by constant star formation from $7$ Gyr
onwards. In this model both the bulge and the nuclear subcomponent
have solar metallicity.  Labelled along the right edges in units of
$\msun/{\rm Gyr}$ are the normalised star formation rates.  }
\label{fig:rcsf}
\end{figure*}

Is the nuclear subcomponent a primordial, undisturbed, spiral-like
disk? In this scenario it is formed continuously from $1$ Gyr onwards
up to the present, and has the same metallicity and abundance as the
bulge. For the purpose of this work, the bulge mass is arbitrary and
set to $1 \msun$.  Line-index predictions from the Big Bang onwards
are presented in Fig.~\ref{fig:csf000-004} for model CSF1 where
$Z_{nuc} = Z_{bulge} =1$, with $SFR$ parameterized in units of
$\msun/{\rm Gyr}$. This model cannot explain simultaneously the
observed weakness of \mean{Mg} and the strength of \fe{5335} (right
hand panel of Fig.~\ref{fig:csf000-004}). Likewise model CSF2, where
$Z_{nuc} = Z_{bulge}=0.4\ \zsun$ cannot explain these parameters.

These models are only approximations because self-enrichment is not
modelled in GISSEL96.  If there is self-enrichment, the metal indices
in Fig.\ref{fig:csf000-004} will have steeper gradients, with \hbeta\
remaining the same due to its weak dependence on metallicity. Whilst
$\Delta \fe{5335}$ is reduced, a very high $SFR$ is required to attain
the observed $\Delta \mean{Mg}$ because many young stars are needed to
dilute the enhanced metal lines in the light of the old stars. As a
rough estimate, model CSF1 is repeated with $Z_{nuc}= 2.5\ \zsun$. We
find that, along the line of sight, the nuclear subcomponent has to be
$\sim 11.2$ times more massive as the bulge. Such a massive disk will
make the isophotes significantly disky, not observed in
NGC~2865. Therefore we conclude the nuclear subcomponent is not a
primordial embedded disk that has been forming stars up to the
present.

\subsection{Recent, continuous star formation (RCSF1,RCSF2)}

Finally, was the nuclear subcomponent formed recently by continuously
feeding gas of constant metallicity into the galaxy centre, for
example, by the return of tidal material after a major spiral-spiral
merger \cite{HM95}?  In this model, the new stars starts
to form at $t_{start}$ and continues constantly up to the present;
their metallicity can be the same or higher than that of bulge.  The
bulge mass is arbitrary and is set to $1\ \msun$. The model
line-indices for model RCSF1 where $Z_{bulge} = Z_{nuc} =1\ \zsun$ are
plotted in Fig~\ref{fig:rcsf}.  The observed $\Delta \hbeta$ \&
$\Delta \mean{Mg}$ are attainable for $SFR =0.05$--0.1 $\msun$/Gyr
within 3 Gyr after $t_{start}$, but the predicted $\Delta \fe{5335}=
-0.75 \ang$ is ruled out at 98\% confidence level. We find that model
RCSF2 is similarly ruled out. The above models are also repeated with
$Z_{bulge}= 0.4\ \zsun$ and $Z_{nuc}= 0.4$ or $1 \zsun$, and we find
that they are also ruled out.  Therefore we conclude the nuclear
subcomponent was not formed by continuous inflow of cold gas with
fixed metallicity on a Gyr timescale.

\subsection{Summary of Population Synthesis Models}

The results of this section are tabulated in Table 3. The observed
line strength indices allow the following models:

\begin{itemize}
\item[] - {\bf Recent burst of star formation with enhanced
metallicity}. A recent burst of star formation works only if the
resulting KDC has a metallicity higher than the galaxy bulge by a
factor of about 2.5--6.3 (models SB3, SB4 \& SB5). The burst duration
can be as long as 0.4 Gyr (model SB8). These models do not fit the
data if the KDC has the same metallicity as the bulge, even if [Mg/Fe]
ratio of the KDC is lower.
\item[] - {\bf Truncated star formation in the KDC, with a difference in
abundance ratio}. In this case the metallicity of the 2 populations are
the same, but the Fe abundance in the KDC is enhanced relative to that
of the bulge by 0.08 dex (model TSF3). The model does not fit the data
without this enhancement.
\end{itemize}

Continuous Star Formation and Recent Continuous Star Formation models
are completely ruled out by the data.

\section{Discussion} \label{sec:discussion}

\subsection*{}
\subsubsection*{A common merger origin for shells and KDC?}

The discovery of a kinematically distinct component at the heart of a
shell galaxy provides additional evidence linking shell systems to
mergers or accretion.  Indeed, the distinct kinematics at the core of
NGC~2865 require formation dynamics decoupled from those of the rest
of the system, and a merger or accretion provides a natural mechanism
for such a process. 

This finding strengthens the link between shells and kinematically
decoupled cores.  \citeN{Forbes92} found that galaxies selected for
their KDC's have a high probability of having shells.  The present
paper has found that a galaxy selected for its shells has a
kinematically distinct core.  In a forthcoming paper (Balcells, Hau \&
Carter, in preparation), we show that NGC~474, a shell lenticular,
shows core kinematic decoupling as well.

The weak interaction model for shells \cite{Thomson91} does not
predict the existence of a kinematically distinct nucleus, as the
shells are produced as a reaction to a grazing incidence passage; the
presence of the rapidly rotating nuclear system in a shell galaxy is
purely coincidental in this framework, unless the core is formed by
mass transfer.  Schiminovich \etal 1995 note the presence of a nearby
gas-rich dwarf, but conclude that it is not related to
NGC~2865. Furthermore because in this model the shells are in an
one-arm density wave, they should follow the brightness of the
underlying elliptical, in conflict with the existence of bright outer
shells, their blue colours, and their chaotic distribution (Fort \etal
1986).  The weak interaction model of decoupled core formation (Hau \&
Thomson 1994) can potentially explain the high $v/\sigma$ of NGC~2865
as due to the {\it spinning up} of the bulge by another galaxy.
However, in this model the nuclear disk is primordial, in conflict
with the observed line indices. It is also unlikely that a passing
galaxy can transfer a large amount of orbital angular momentum over a
period longer than 0.5 Gyr without being captured or substantially
disrupted, as NGC~2865 has an extended massive dark halo (Schiminovich
\etal 1995). Thus a purely interaction induced origin for the shells
and KDC in NGC~2865 is ruled out.

\subsubsection*{A gas rich merger and starburst origin for the KDC?}

The presence of a young population in the galaxy core is evidence for
a late merger event.  Not only is the population young, but it may be
formed in a burst, a typical result of a spiral-spiral (SS) merger, or
of the accretion of a gas-rich object by an elliptical.  The distinct
core can be formed by a starburst with metallicity $2.5$--$6$ times that
of the underlying galaxy, with $0.13 \le f_b \le 0.91$ and a short
($\ale 0.4$ Gyr) duration.  The best fit age, which ranges from 0.4 to
1.7 Gyr, agrees with the age determined by \citeN{BA87}.
It depends on $f_b$ and metallicity, with more recent bursts favored
for low $f_b$ or high metallicity, and vice versa. 

The range of metallicity
enhancement required is in good agreement with those expected by
theory (\citeNP{fva-gerhard94}, \citeNP{mihos-hernquist94apj427}). By
considering a 1-zone closed boxed model, Fritze-v. Alvensleben \&
Gerhard find that immediately ($\ale 0.1$ Gyr) after a collision of a
spiral pair at 12 Gyr which consumed all their remaining gas, the
global metallicity enhancement factors are about 1.5, 1.9 \& 7.5 for
Sa, Sb and Sc pairs respectively (Figs 12a \& 12b of
Fritze-v. Alvensleben \& Gerhard 1994). The metallicity of NGC~2865 is
roughly solar, similar to galaxies earlier than Sc.  Taking both
metallicity enhancement and absolute metallicity into account, if
NGC~2865 is the remnant of the merger of 2 spirals of equal Hubble
type, they are most likely to be of type between Sb \& Sc (although
see later discussion).

The metallicity observed in the peculiar core of NGC~2865 is higher,
by large factors, than that of early type spiral bulges \cite{BP94} and that of HII regions of disks \cite{VE92}, hence self-enrichment must have taken place, suggesting
a burst duration longer than SN II feedback timescale.  SPH
simulations of spiral-spiral (SS) mergers which include star-formation
\cite{mihos-hernquist96apj464} show that galaxy structure plays a
dominant role in regulating starburst activities.  If the progenitors
have massive central bulges, gas inflows are strongest at the final
stage of merging, and the resulting starburst is intense but
short-lived ($\sim$0.02--0.03 Gyr), whereas if the progenitors are
bulgeless (or have small bulges), the starburst is strongest after the
first passage of the galaxies, with successive episodes of star
formation after each passage, and the final merger consumes the rest
of the gas. Therefore successive generations of starbursts have higher
and higher metallicity. Furthermore, the burst durations in bulgeless
encounters are much longer ($\sim 0.15$ Gyr), thus allowing more time
for self-enrichment in each starburst. Therefore a late type spiral
progenitor with a small bulge is favored over an early type spiral
with a large bulge. This will also be consistent with the shell colour
indices which Fort \etal (1986) found to resemble those of an Sb or Sc
galaxy.

Although the above discussions are based on simulations of SS mergers,
because gas inflow is driven by the tidal torque of the galaxy the gas
resides in, the general picture above will probably be true for a
elliptical-spiral (Es or ES) merger too. 

\subsubsection*{The identity and mass ratio of the progenitors}

Does our data allow us to clarify the discrepant views on the type of
merger which formed NGC~2865 (\S~1) ?  If the shells contribute $\sim
11-22$\% of the total light of the galaxy \cite{fort-etal86}, then
taking into account that only a small fraction of a merged galaxy ends
up in visible shells, the fraction of the total system mass intially
in dynamically cold components cannot be much lower than 0.5.  Hence,
if one galaxy was an elliptical, the galaxy mass ratio must have been
close to unity.  This argues against the minor merger scenario
proposed by Bettoni (1992).  Alternatively, both progenitors could
have been disk galaxies.  In the disk-disk case, we can constrain the
mass ratio to below a value around 3:1 or 5:1, as higher mass ratios
do not destroy the disk of the larger galaxy \cite{barnes1996fogh}.
The large amount of rotational support ($v/\sigma \sim 1.2$), one of
the highest known for an elliptical, could be a relic of the disk
rotation of the larger precursor in an intermediate mass-ratio merger.
It could also result from a good alignment of the two spins in an 1:1
SS merger.  The good alignment between stellar and gas kinematics
could be a relic of disk rotation in an intermediate mass merger of
two spirals, or the result of an organizing effect of the merger in an
equal mass SS merger.

\subsubsection*{Discrepancy between age of shells and the KDC--
against a phase-wrapping origin of shells?}

The ages estimated for the NGC~2865 starburst are higher than typical
ages for shell systems. Using phase wrapping theory, \citeN{HQ87} derive an age of $\sim 5 \times 10^8\,h_{100}^{-1}$ yr for
NGC~3923, $\sim 10^9\,h_{100}^{-1}$ yr for NGC~1344 and $\sim 2 \times
10^8\,h_{100}^{-1}$ yr for NGC~3051. For a given mass model, the time
since the merger scales as
\begin{equation}
t_{merger} \approx  (n+0.5) P(d_n)
\label{eqn:d1}
\end{equation}
(formula [5.3] of \citeN{HQ87}), where $n$ is the
number of shells and $P(d_n)$ is the time between turning points (half
the orbital period) at the location of shell $n$ (the innermost
shell).  Large galaxy sizes and a high number of shells translate into
a long merger age.  Notwithstanding that the chaotic nature of the
shells in NGC~2865 precludes a rigorous application of phase-wrapping
theory, we apply this formula directly, seeking an order-of-magnitude
result.  We further approximate
\begin{equation}
P(d_n) \approx 2 d_n / \sigma, 		
\label{eqn:d2}
\end{equation}
which should be acceptable for our purposes if $d_n \approx r_e$.  We
take $n=9$, i.e. the 7 shells identified by Fort et al. (1986) plus
the two outer loops seen in Fig.~1$b$ of \citeN{schiminovich-etal95};
$d_n$ = $20\asec$ = 2.5\,$h_{100}^{-1}$ kpc, $r_e = 16.6\asec =
2.1\,h_{100}^{-1}$ kpc \cite{jorgensen92}, and $\sigma = 200\ \kms$
(this paper).  We obtain
\begin{equation}
t_{merger} \approx 2.4 \times 10^8\,h_{100}^{-1} {\rm yr} 		
\label{eqn:d3}
\end{equation}
Hence, phase-wrapping yields $t_{merger}$ an order of magnitude below
the burst age derived from population synthesis.  Uncertainties in the
true number of shells, or in the number of periods completed by stars
in the outermost shell, can at most introduce a relative error of a
few tenths in $t_{merger}$, as they contribute additive terms of order
1 to $n$ in formula ~\ref{eqn:d1}.  Approximation (\ref{eqn:d2}) for
$P(d_n)$, when applied to NGC~3923, gives $t_{merger}$ in accord with
the full model calculation of \citeN{HQ87} to within
25\%.  Thus, it is unlikely that the approximations in the calculation
can account for the discrepancy between the phase-wrapping merger age
and the age derived for the stellar population. In theory, the age of
the KDC can be shortened by raising its metallicity to higher than 6.3
times that of the the bulge.  More likely, this discrepancy indicates
that the shell system did not form by phase-wrapping of a
dynamically-cold, accreted satellite, and that the phase wrapping
theory is not applicable to NGC~2865, either because of dynamical
friction or because a major, rather than a minor, merger took
place. Alternative mechanisms include spatial wrapping or phase
wrapping in a SS merger remnant.  These are supported by Fort \etal
(1986), who remarked on the similarity between NGC~2865 and
parabolic-encounter models (e.g. model 8 in Hernquist \& Quinn 1987),
where shells are formed by spatial wrapping over several orbits, whose
formation timescale could be as long as 4 Gyr.

\subsubsection*{Truncated star formation?}
Another way out of the age discrepancy problem is to form the KDC by
accreting stars from a small star-forming disk galaxy into a giant
elliptical, during which the star formation in the former is
truncated. The two progenitor galaxies have the same Mg abundance, but
the accreted population has a relatively higher Fe abundance. A high
$f_b$ of $0.5 \ale f_b \ale 0.99$ is required.  The age estimates of
$0.1$ -- $0.4$ Gyr is in better agreement with the age of the
phase-wrapped shells. The physical processes involved in this model
are contrived.  An accreted stellar system can create a central
concentration in the remnant via dynamical friction only
\cite{BQ90}, and a small spiral is likely to tidally disrupt
before reaching the center of the elliptical. However, it is
conceivable that a relatively gas-poor and small spiral sinks
undisrupted thanks to the self gravity of its dark halo, forming the
kinematically distinct core which dominates the nuclear light, whilst
tidally-ejecting parts of its disk to become phase-wrapped shells. We
note that if only the bulge remains bound, however, there will be a
problem matching the age and metallicity observed in the center, as
bulges of small spirals are not metal rich \cite{BP94}.

It is also unclear how to truncate the star formation in a
star-forming galaxy during a merger. Below, we outline briefly the
possibilities--- quantitative discussion is beyond the scope of this
work. It may be possible that during the merger process, young stars
are formed by tidal shocking of gas along transient spiral arms, but
they are not confined to the nucleus, where the old spiral disk
dominates the light.  Another possibility is the evaporation of the
spiral HI by the hot ISM of the elliptical, which may effectively and
cleanly truncate the star formation, but may only work for a very
small accreted galaxy.  We note that SPH models of elliptical-spiral
(Es) mergers (e.g. \citeNP{HW92}) do not model the hot
ISM in the elliptical.  Other possibilities are ram-pressure stripping
of HI by the hot elliptical ISM, or the stripping of the remaining HI
by the winds of the first supernovae. Ram-pressure stripping could
possibly explain the displacement of HI outside the shells reported by
\shortciteN{schiminovich-etal95}.

Finally, given that both a starburst and a truncated star-formation
scenario explain the line indices, it is possible that both processes
played a role in NGC~2865.  Indeed, a natural star formation history
for a gas-rich galaxy may be that of forming stars throughout its
life, and at some point having its remaining gas consumed in a
starburst after a collision.

\subsubsection*{Can the age difference between the shells and the 
KDC reflect the merging time?}

A further possibility is that the shells and the decoupled core are
not formed at the same time. This is quite plausible, as phase-wrapped
shells are formed at the final stage of a merger, whereas starburst
activities are regulated by the internal structure of the spiral, and
maybe strongest after the first passage and well before the final
merger for a bulgeless spiral
\cite{mihos-hernquist96apj464}. Therefore if the shells are indeed
phase-wrapped, the age difference between them and the KDC may reflect
the merging time, and further favor a late-type spiral progenitor.

\subsubsection*{A comment on Mg enrichment and the adopted IMF}

Throughout this paper, a Scalo IMF has been adopted. A proper
investigation on the choice of IMF is beyond the scope of this work,
and is further hampered by the fixed abundance ratios in the models,
and the lack of knowledge on self-enrichment of individual
elements. Starbursts are suspected to produce more massive stars,
which are thought to feedback more Mg than Fe into the ISM through
type II supernovae. We find in \S~\ref{sec:mgfevst} that the \mean{Mg}
of a burst population is weaker than the \mean{Fe} before 1.7 Gyr, due
to a temperature effect in the atmospheres of stars at the MS turnoff.
This helps to hide any enhanced Mg abundance during this
time. Therefore, unless the Mg is much more enriched than the Fe
during the starburst, our results are probably insensitive to the Mg
enrichment. This work would benefit from a better understanding of the
role SN feedback, gas dynamics and star-formation play in gas-rich
mergers.

\subsubsection*{The ``bulge+burst'' assumption, and the departure from the 
$Mg-\sigma$ relationship}
\begin{figure}
\psfig{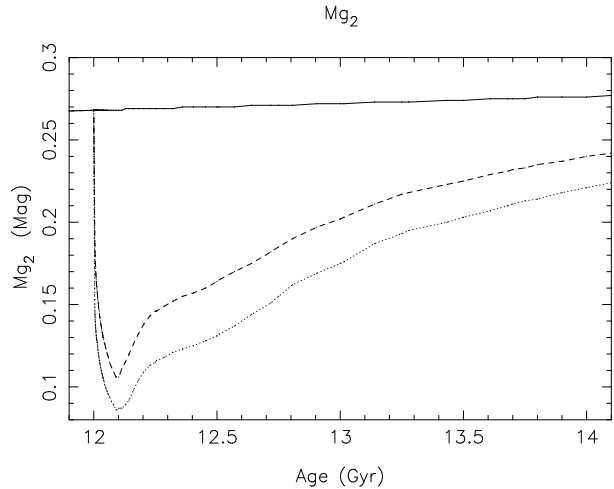}
\caption[]{The effect of light contamination by young stars on an old
bulge population. The bulge is formed between $0$ \& $1$~Gyr with
solar metallicity. Its \mg2 is indicated by the solid line. Added to this
population are young stars of the same metallicity formed between 12.0
and 12.1 Gyr. The dashed line shows the \mg2 for the combined light,
where the young stars contribute $9\%$ of the total mass. The dotted
line shows the same for a $17\%$ contribution by young stars. }
\label{fig:mg2fundamental}
\end{figure}

The very ``old'' nature of the underlying population is an assumption
of the synthesis models.  Outside the young core ($r \geq 10''$)
\hbeta\ $\approx 2 \ang$ (Fig.~\ref{fig:n2865indices}).  A roughly
10\% pollution by stars of order 1 Gyr old onto an underlying 12 Gyr
population results in such an \hbeta\ value
\cite{dejong-davies97}. This will also be consistent with the blue
shells which contribute $\sim 11-22\%$ of the total luminosity.  In
\S~\ref{sec:indices} we find that the \mg2 over the galaxy is about
0.08 mag below that expected from the $Mg-\sigma$ relationship.  This
has also been noted independently by \citeN{jorgensen97}. Is this
scatter from the fundamental plane due to light contamination by a
young population? To investigate this, we add young stars to an old
solar-metallicity population, and the model \mg2 are plotted in
Fig.~\ref{fig:mg2fundamental}. We find that a roughly 10\%
contamination by 1 Gyr old stars can indeed result in the magnitude of
scatter observed. Therefore the young population is not only confined
to the nucleus, but a fraction is also mixed into the underlying
elliptical population. This makes the galaxy appear younger or has a
lower metallicity. In light of this, the models in table 3 with a
solar metallicity bulge are probably more relevant than those with
sub-solar metallicity.

NGC~2865 shows that gas-rich, shell-forming mergers can be a source of
scatter from the $Mg-\sigma$ relationship.  The ``bulge+burst''
assumption is an approximation which helps to keep the models
simple. But because the light contamination by the young stars on the
bulge is small, the conclusions from population synthesis are unlikely
to be significantly affected.

\subsubsection*{KDC formation issues}

Our discovery of a KDC in a shell elliptical galaxy, with associated
young stellar population, fits well in the framework of gas rich
spiral merger simulations (\citeNP{HB91},
\citeNP{HW92}). The fact that the light profile of
NGC~2865 is well fit by a $r^{\frac{1}{4}}$ law, with a kinematically
distinct, compact and young component embedded in a dynamically hot
system, is consistent with the segregation of stars and gas seen in
these simulations.

We demonstrated that despite the age and metallicity degeneracy of the
underlying elliptical, physical parameters such as the age and
metallicity of the kinematic subcomponent in a recent merger can be
constrained. In addition we discovered that, between about 0.4--1.8
Gyr after a metallicity-enhanced starburst, the \mean{Fe} of stars at
the MS turnoff is strong enough to compensate for dilution effects,
whilst \mean{Mg} is weaker. As a result, the abundance ratio of the
total light appears to be modified.  This is purely a temperature
effect in the atmospheres of stars at the MS turnoff.  As a
consequence of the weaker Mg index in the burst population, Mg
enrichment by a starburst may be partially hidden during approximately
1.8 Gyr.

In a sample of galaxies chosen subjectively by their visual appearance
to represent the time sequence of SS mergers, NGC~2865 was ranked as
one of the oldest remnant identifiable by its morphological features
\cite{KW95}.  That ranking was based on the basis of the
shells and deviations from elliptical shape.  The dating of such a
galaxy by morphology is difficult because tidal tail remnants
disappear after a few crossing times. It is also unlikely that visual
appearance alone can provide enough evidence for a SS merger origin
for a remnant like NGC~2865, unlike NGC~7252 \cite{Schweizer82} or
NGC~3656 \cite{balcells2tails97}.  This work demonstrates that an
older merger remnant may still be dated by its stellar populations,
even when the tidal tail has virtually disappeared.
 
NGC~2865 has the only KDC with depressed Mg indices-- most, if not
all, decoupled cores have enhanced Mg, in common with most giant
ellipticals \cite{BS92}. Pointing to the different
[Mg/Fe] abundance ratios between spirals and giant ellipticals, and
the fact that KDCs tend to reside in the latter, \citeN{bender96iau}
favored the view that KDCs are metallicity-enriched remnants of early
hierarchical mergers rather than remnants of present-day S-S mergers.
We find that NGC~2865 could be the renmant of a recent S-S or S-E
merger. What would the metallicity of the decoupled core be in the
next few Gyrs?  If it was formed by a metallicity enhanced starburst,
then its metal line-indices would eventually be higher than those of
the underlying elliptical. If the IMF is top-heavy and the
star-formation timescale is short, there would be an additional Mg
enhancement observable after $\sim 1.9$ Gyr \cite{MB90}.
Thus a recent starburst origin will fit the overall scheme of KDC
formation.  NGC~2865 is an example of a young KDC formed in a recent
merger. Unlike for instance NGC 7626 \cite{BC93} NGC
2865 has symmetric, regular velocity field. It is unclear what
proportion of young KDCs do have a regular velocity field, and
therefore how quickly the velocity field relaxes.

\section*{Acknowledgments} 
We are grateful to Hans-Walter Rix for providing us his program for
the line-profile analysis, and to Gustavo Bruzual \& St\'ephane
Charlot for their spectral synthesis code. We thank Alfonso
Aragon-Salamanca, Terry Bridges, Reynier Peletier, Bianca Poggianti \&
Roberto Terlevich for encouragements and valuable discussions.  GKTH
acknowledges a Postgraduate Research Studentship from PPARC and an
Isaac Newton Studentship from Cambridge University.

\appendix

\section{Recoverability of kinematic parameters in the presence of noise}
\label{sec:noisetest}
\begin{figure}
  \centerline{\psfig{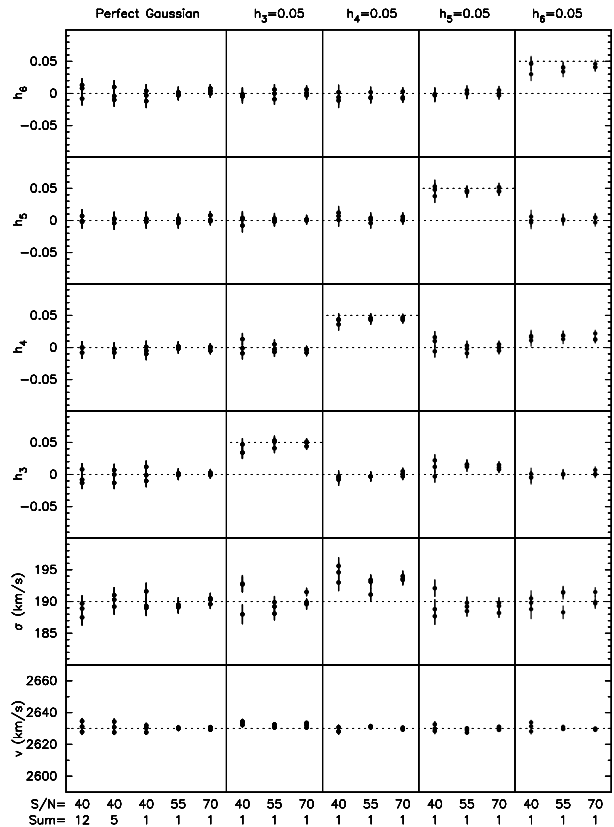}}
\caption[]{The effect of noise to the recovery of the parameters. The
composite stellar template at the nucleus of NGC~2865 is convolved
with different model LOSVDs to make mock galaxy spectra, and added to
them Poisson noise and blank sky spectra. The kinematic parameters are
recovered 3 times, each with a different noise seed and sections of
blank sky. The leftmost panel corresponds to results for a Gaussian
LOSVD, and the next 4 panels correspond to results with a signal of
$0.05$ in each of $\h3$, $\h4$, $\h5$ \& $\h6$ respectively. The model
values are represented by the dotted lines, and the recovered
parameter the solid dots. In each panel, results are shown for $S/N$
of 40, 55 and 70, labelled along the bottom with the number of rows of
sky summed. For a Gaussian LOSVD, 5 and 12 rows of sky spectra are
added to the artificial spectrum to mimmick the points farthest from
the galaxy nucleus. }
\label{fig:noisetest2865}
\end{figure}

The sensitivity of the kinematic parameters to noise is investigated
by simulating the effect of both Poisson shot noise and residuals from
imperfect sky subtraction.  The artificial data is generated by
convolving a stellar template with Gaussian and non-Gaussian LOSVDs of
$cz=2630\,\kms$ and $\sigma=190\,\kms$, scaled to different continuum
levels, and added Poisson noise to make spectra with $S/N$ = $40$,
$55$ and $70$, typical of our data. The spectra at the nucleus of
NGC~2865 have $S/N \sim 70$, whilst those at intermediate distances
($\sim 2\asec$--$3\asec$) have $S/N \sim 40$.  The effect of imperfect
sky subtraction is mimicked by adding rows of real `sky' spectra
extracted sufficiently away from the galaxy. For a Gaussian LOSVD, up
to 12 rows are added to mimic the worst case of sky contamination
where heavy summation is required to attain the desired $S/N$
($\age40$) typical of the outermost points.  Recovery of non-Gaussian
LOSVDs is tested by setting one Hermite moment to $0.05$ whilst
keeping the others zero.  The kinematic parameters are recovered 3
times using {\tt kinematics}, each with different Poisson noise seeds
and sections of blank sky.

The results of the investigation are plotted in
Fig.~\ref{fig:noisetest2865}.  The mean velocity $v$ can be determined
to within $\sim 3\,\kms$ at the galaxy nucleus, and $\sim 20\,\kms$ at
the outermost points. It is insensitive to asymmetric deviations (\h4
\& \h6), with only a small positive offset when $\h3$ is introduced to
the LOSVD, and no significant offset when $\h5$ is introduced.

If the LOSVD is Gaussian, the velocity dispersion $\sigma$ is at best
determined to within $\sim 3\,\kms$ at the nucleus, and to $\sim 5\
\kms$ at intermediate distances. Non-Gaussian LOSVDs increase the
scatter further, with the introduction of $0.05$ in each Hermite
moment increasing the scatter in $\sigma$ by about $2\,\kms$ at the
nucleus, and about $5\,\kms$ at intermediate distances. In addition,
an $\h4$ of $0.05$ introduces a systematic displacement of $\sim +4\
\kms$ in $\sigma$.

In general, the Hermite moments \h3, \h4, \h5 and \h6 can be
recovered with reasonable success, although with decreasing $S/N$,
the signal cannot be fully recovered.  Furthermore, signals in
higher-order Hermite moments tend to leak into the moment $2$ orders
below. This effect is independent to $S/N$, and is probably a
consequence of the finite resolution of the data.  A Hermite moment
\h3 of $0.05$ can be recovered to within $0.01$ at the galaxy nucleus,
and to within $0.02$ at intermediate distances.  An \h5 of $0.05$
introduces a systematic displacement in \h3 of about $+ 0.01$.  Apart
from a small positive displacement in velocity, no systematic
displacement is introduced in the other parameters when $\h3$ is set to
$0.05$. An \h4 of $0.05$ can be recovered to within $0.01$ at the
nucleus, and within about $0.02$ at intermediate distances. Non-zero
values in other Hermite moments tend to increase the scatter in \h4
slightly.  In addition, a systematic displacement of $\sim +0.02$ in
\h4 is introduced when \h6 is $0.05$.

The higher moments \h5 and \h6 can be recovered reasonably well, with
a scatter of $0.01$ at the nucleus and $0.03$ for the intermediate
points. The \h6 cannot be recovered fully, and is systematically
lowered by about $0.01$. 

There appears to be little problems associated with skyline-residuals
from imperfect sky-subtraction.  Even when 12 rows of spectra are
combined to attain the desirable $S/N$, the scatter in the recovered
parameters are increased only slightly compared with those derived
from high-quality spectra, and only with $v$ most affected.

The results from these exercises show that Hermite moments up to \h6
can be recovered reasonable well for the points inside $\sim 3
\asec$. The signals in the lower Hermite moments may be affected if
there is significant signal in the Hermite moment 2 orders up.  The
parameter least sensitive to noise is the mean velocity, whilst the
velocity dispersion $\sigma$ is affected by lowering the $S/N$ or by
increasing the power in the Hermite moments.
 
\section{Dependence of the kinematic parameters on the choice of templates} 
\label{sec:template-tests}

\begin{figure}
  \centerline{\psfig{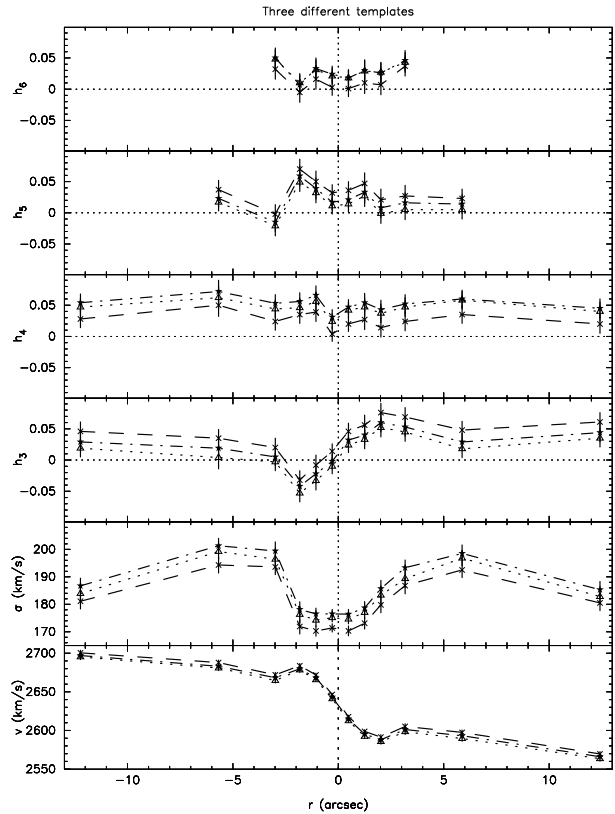}}
\caption[]{The Gauss-Hermite parameters \h6, \h5, \h4, \h3, $\sigma$
and $v$ recovered separately using the 3 available stellar templates,
showing that their zero-points depend on the choice of template, but
not their radial profile.}
\label{fig:templates-mismatch}
\end{figure}

Apart from noise, the kinematic parameters are sensitive to template
mismatch, which could be caused by a difference in stellar type
between the galaxy population and the template, or by a difference in
metallicity abundance. The former can be somewhat minimised by
utilisation of an optimal template, taken as the best linear
combination of the stellar templates (spanning a range of stellar
type) that minimises the $\chi^2$, after an initial estimation of the
LOSVD. In theory, the latter could be minimised too if the abundances
of the template stars span a range that brackets that of the galaxy.
The Lick indices of the template stars span the range 0.015--0.096 mag
in \mg1, 0.090--0.220 mag in \mg2, 1.52--3.43 $\ang$ in \mgb,
3.07--4.30 $\ang$ in \fe{5270} and 2.74--3.92 in \fe{5335}. Thus the
\mg1, \mg2 and \fe{5335} are well matched, and the \mgb\ and \fe{5270}
marginally matched. Thus template mismatch due to metal abundance
difference is unlikely to be a problem.  We note if absorption lines
of the stars are much weaker than that of the galaxy, then the \h4 may
be artificially enhanced as the model spectrum is made more ``pointy''
to match the galaxy spectrum.

For the purpose of quantifying the sensitivity of the Gauss-Hermite
parameters to template mismatch, they are derived separately with each
of the 3 available stellar templates. The results are plotted in
Fig.~\ref{fig:templates-mismatch}.  The parameters's zero points vary
with the choice of template but their radial profiles do not.  The
extreme differences are $6\,\kms$ for both $v$ \& $\sigma$, and
$0.026$, $0.026$, $0.020$ and $0.018$ for \h3, \h4, \h5 \& \h6
respectively. These should be regarded as the maximum zero point
offsets, because the employment of an optimal template ensures that
template mismatch is minimized. We conclude that the radial profiles
of the Gauss-Hermite moments, apart from \h4, cannot be explained by
radial template mismatch even at the most extreme. The lowered point
in \h4 at $r=0.3 \asec$ is reproduceable by extreme template mismatch.

In the final analysis (see fig.~\ref{fig:n2865hermplot} of main text),
the zero-point of $\sim +0.02$ in \h3 is explainable by mismatch, and
likewise \h5 and \h6.  The absolute zero-point of the \h4 is related
to the radial anisotropy of the orbital population.  The constant \h4
offset of $\sim +0.045$ along the major axis is larger than the
extreme difference of $0.026$ introduceable by the 3 templates. Even
taking into account the result from the previous section where $\h6
\sim 0.02$ {\it may} introduce an additional offset of $\sim 0.01$
into the \h4, its offset is still marginally significant. 


\begin{table*}
\label{tab:n2865maj}
\centering
\caption{Measured parameters along the major axis of NGC~2865.}
\vspace{0.5cm} 
\begin{tabular}{lllllllllllll}
\hline
$r (\asec)$ & $v $ & $err_{v} $ & $\sigma $ & $err_{\sigma}$ & $h_3$ & $err_{h_3}$ & $h_4$ & $err_{h_4} $ & $h_5$ & $err_{h_5}$ & $h_6$ & $err_{h_6}$  \\
\\
$-12.2$& 2691.3& 3.4& 187.7& 3.1& 0.035& 0.016& 0.021& 0.015& - & -& - & -\\
$-5.7$ & 2681.0& 3.6& 199.1& 2.9& 0.008& 0.015& 0.041& 0.014& 0.036& 0.015& - & -\\
$-3.0$ & 2665.1& 3.6& 195.5& 3.5& $-0.005$& 0.016& 0.032& 0.015& $-0.009$& 0.016& 0.040& 0.016\\
$-1.8$ & 2675.1& 3.2& 176.0& 2.7& $-0.037$& 0.015& 0.044& 0.015& 0.054& 0.016& 0.000& 0.016\\
$-1.1$ & 2664.4& 2.2& 178.2& 2.1& $-0.011$& 0.016& 0.049& 0.015& 0.039& 0.017& 0.027& 0.017\\
$-0.3$ & 2642.0& 1.9& 177.0& 1.8& $-0.001$& 0.012& 0.012& 0.012& 0.017& 0.013& 0.014& 0.013\\
0.5    & 2614.0& 1.9& 175.6& 1.8& 0.028& 0.013& 0.034& 0.012& 0.016& 0.014& 0.016& 0.014\\
1.3    & 2595.7& 2.6& 180.2& 2.1& 0.041& 0.016& 0.028& 0.016& 0.020& 0.017& 0.018& 0.017\\
2.0    & 2588.1& 3.7& 187.1& 3.2& 0.058& 0.016& 0.025& 0.016& 0.012& 0.017& 0.014& 0.017\\
3.2    & 2599.3& 3.8& 192.3& 3.5& 0.059& 0.016& 0.022& 0.016& 0.018& 0.017& 0.046& 0.017\\
5.9    & 2588.0& 3.2& 197.6& 3.3& 0.024& 0.015& 0.030& 0.015& 0.012& 0.016& - & -\\
12.4   & 2564.8& 3.6& 183.4& 3.0& 0.038& 0.016& 0.011& 0.015& - & -& - & -\\
\hline
\end{tabular}
\end{table*}

\begin{table*}
\label{tab:n2865min}
\centering
\caption{Measured parameters along the minor axis of NGC~2865.}
\vspace{0.5cm} 
\begin{tabular}{llllllllllllll}
\hline
$r (\asec)$ & $v $ & $err_{v} $ & $\sigma $ & $err_{\sigma}$ & $h_3$ & $err_{h_3}$ & $h_4$ & $err_{h_4} $ & $h_5$ & $err_{h_5}$ & $h_6$ & $err_{h_6}$ \\
\\
$-4.8$&2632.4 & 3.7& 197.0& 3.6& 0.017& 0.017& $-0.007$& 0.017& - & -& - & -\\
$-2.5$&2640.5 & 3.1& 205.7& 3.1& 0.020& 0.014& 0.024& 0.014& 0.011& 0.014& - & -\\
$-1.4$&2632.7 & 2.6& 184.8& 2.4& 0.010& 0.017& 0.036& 0.016& 0.037& 0.018& 0.039& 0.014\\
$-0.6$&2629.1 & 2.4& 185.5& 1.5& $-0.008$& 0.013& 0.016& 0.013& 0.029& 0.013& 0.014& 0.013\\
0.2   &2630.9 & 1.7& 192.8& 1.6& 0.002& 0.012& 0.016& 0.012& 0.019& 0.012& 0.023& 0.013\\
1.0   &2638.7 & 2.4& 185.5& 2.2& $-0.033$& 0.016& 0.026& 0.015& 0.023& 0.017& 0.031& 0.017\\
1.7   &2629.9 & 3.5& 190.6& 3.2& 0.016& 0.016& 0.035& 0.015& 0.018& 0.017& 0.050& 0.017\\
2.9   &2629.9 & 3.6& 188.3& 3.5& 0.018& 0.017& 0.046& 0.016& 0.067& 0.018& - & -\\
6.0   &2624.1 & 4.0& 202.2& 3.5& 0.008& 0.016& 0.052& 0.016& - & -& - & -\\
12.5  &2615.2 & 5.1& 175.3& 4.4& 0.012& 0.022& - & -& - & -& - & -\\
\hline
\end{tabular}
\end{table*}


\begin{table*}
\label{indices-major.tab}
\centering
\caption{Measured line indices and their errors along the major axis of NGC~2865.}
\vspace{0.5cm} 
\begin{tabular}{lllllllllllll}
\hline
$r (\asec)$&$H_{\beta}$&$err$&$Mg_1$&$err$&$Mg_2$&$err$& $Mg\ b$ & $err$ & $Fe_{5270}$ & $err$ & $Fe_{5335}$ & $err$ \\ \\
  -14.92  &  -    &   -   & 0.070  & 0.016  &   0.180  &  0.020 &  4.07  & 0.72  &    -  &    - &     -  & -    \\                  
  -14.15  &  2.57 & 0.37  & 0.070  & 0.016  &   0.184  &  0.020 &  2.46  & 0.73  & 2.72  & 0.43 &  3.41  & 0.41 \\
  -13.38  & -     &   -   & 0.071  & 0.015  &   0.208  &  0.019 &  3.33  & 0.69  &    -  &    - &     -  &  -   \\  
  -12.61  & -     &   -   & 0.067  & 0.014  &   0.198  &  0.018 &  2.35  & 0.68  &    -  &    - &     -  &  -   \\  
  -11.84  &  2.41 & 0.33  & 0.074  & 0.014  &   0.188  &  0.018 &  4.52  & 0.61  & 2.91  & 0.38 &  2.39  & 0.38 \\ 
  -11.07  & -     &   -   & 0.067  & 0.013  &   0.194  &  0.017 &  3.42  & 0.61  &    -  &    - &     -  &  -   \\  
  -10.30  &  3.46 & 0.51  & 0.068  & 0.013  &   0.190  &  0.016 &  3.88  & 0.57  & 2.23  & 0.61 &  2.84  & 0.60 \\ 
   -9.53  &  1.74 & 0.50  & 0.076  & 0.012  &   0.194  &  0.015 &  3.63  & 0.55  & 2.67  & 0.57 &  2.68  & 0.57 \\ 
   -8.76  &  2.82 & 0.47  & 0.074  & 0.011  &   0.188  &  0.014 &  3.60  & 0.51  & 2.41  & 0.54 &  2.53  & 0.53 \\ 
   -7.99  &  1.77 & 0.45  & 0.066  & 0.011  &   0.193  &  0.013 &  4.24  & 0.48  & 3.06  & 0.51 &  2.89  & 0.50 \\ 
   -7.22  &  2.07 & 0.42  & 0.078  & 0.010  &   0.199  &  0.013 &  3.89  & 0.45  & 2.60  & 0.48 &  3.16  & 0.47 \\ 
   -6.45  &  2.23 & 0.39  & 0.081  & 0.009  &   0.194  &  0.012 &  3.85  & 0.42  & 2.27  & 0.45 &  2.79  & 0.44 \\ 
   -5.68  &  2.45 & 0.37  & 0.077  & 0.009  &   0.204  &  0.011 &  3.58  & 0.39  & 2.87  & 0.41 &  2.54  & 0.41 \\ 
   -4.91  &  2.50 & 0.33  & 0.082  & 0.008  &   0.199  &  0.010 &  3.56  & 0.36  & 2.52  & 0.38 &  2.60  & 0.37 \\ 
   -4.14  &  2.74 & 0.30  & 0.085  & 0.007  &   0.211  &  0.009 &  3.93  & 0.32  & 2.86  & 0.34 &  2.67  & 0.34 \\ 
   -3.37  &  2.76 & 0.27  & 0.084  & 0.006  &   0.208  &  0.008 &  3.82  & 0.29  & 3.26  & 0.30 &  2.76  & 0.30 \\ 
   -2.60  &  2.72 & 0.22  & 0.088  & 0.005  &   0.198  &  0.006 &  3.77  & 0.23  & 2.65  & 0.25 &  2.90  & 0.24 \\ 
   -1.83  &  2.92 & 0.18  & 0.094  & 0.004  &   0.199  &  0.005 &  3.38  & 0.18  & 2.71  & 0.19 &  2.85  & 0.18 \\ 
   -1.06  &  3.50 & 0.15  & 0.082  & 0.003  &   0.189  &  0.004 &  3.28  & 0.13  & 2.83  & 0.13 &  2.68  & 0.13 \\ 
   -0.29  &  3.44 & 0.13  & 0.058  & 0.002  &   0.170  &  0.003 &  3.36  & 0.11  & 2.74  & 0.11 &  2.85  & 0.11 \\ 
    0.48  &  3.45 & 0.13  & 0.037  & 0.002  &   0.153  &  0.003 &  3.31  & 0.11  & 2.82  & 0.11 &  2.76  & 0.11 \\ 
    1.25  &  3.33 & 0.14  & 0.058  & 0.003  &   0.172  &  0.004 &  3.41  & 0.14  & 2.87  & 0.14 &  2.71  & 0.14 \\ 
    2.02  &  3.29 & 0.18  & 0.076  & 0.004  &   0.192  &  0.005 &  3.51  & 0.19  & 2.87  & 0.20 &  2.75  & 0.20 \\ 
    2.79  &  2.81 & 0.22  & 0.079  & 0.005  &   0.194  &  0.007 &  3.65  & 0.24  & 2.95  & 0.25 &  2.72  & 0.25 \\ 
    3.56  &  2.73 & 0.26  & 0.075  & 0.006  &   0.197  &  0.008 &  3.93  & 0.29  & 2.61  & 0.31 &  2.73  & 0.30 \\ 
    4.33  &  2.94 & 0.30  & 0.069  & 0.007  &   0.193  &  0.009 &  3.79  & 0.33  & 2.81  & 0.35 &  2.59  & 0.34 \\ 
    5.10  &  3.20 & 0.33  & 0.063  & 0.008  &   0.202  &  0.010 &  4.15  & 0.36  & 2.84  & 0.38 &  2.82  & 0.38 \\ 
    5.87  &  2.18 & 0.36  & 0.068  & 0.009  &   0.199  &  0.011 &  3.99  & 0.39  & 2.72  & 0.42 &  2.52  & 0.42 \\ 
    6.64  &  2.33 & 0.39  & 0.069  & 0.009  &   0.173  &  0.012 &  3.71  & 0.43  & 2.63  & 0.45 &  3.15  & 0.44 \\ 
    7.41  &  2.18 & 0.42  & 0.083  & 0.010  &   0.192  &  0.013 &  3.65  & 0.46  & 2.18  & 0.49 &  2.77  & 0.48 \\ 
    8.18  &  2.41 & 0.45  & 0.073  & 0.011  &   0.196  &  0.014 &  4.27  & 0.48  & 2.18  & 0.53 &  2.66  & 0.51 \\ 
    8.95  &  2.53 & 0.47  & 0.067  & 0.012  &   0.187  &  0.014 &  4.09  & 0.51  & 2.99  & 0.55 &  2.01  & 0.56 \\ 
    9.72  &  1.89 & 0.51  & 0.077  & 0.012  &   0.184  &  0.015 &  3.68  & 0.55  & 2.40  & 0.59 &  2.76  & 0.57 \\ 
   10.49  &  1.65 & 0.54  & 0.060  & 0.013  &   0.194  &  0.016 &  3.34  & 0.59  & 3.03  & 0.60 &  3.73  & 0.59 \\ 
   11.26  & -     &   -   & 0.051  & 0.013  &   0.177  &  0.017 &  4.23  & 0.59  &    -  &    - &     -  &  -   \\  
   12.03  &  1.77 & 0.34  & 0.062  & 0.014  &   0.187  &  0.018 &  3.45  & 0.63  & 2.28  & 0.38 &  2.10  & 0.38 \\ 
   12.80  & -     &   -   & 0.058  & 0.015  &   0.161  &  0.018 &  3.63  & 0.64  &    -  &    - &     -  &  -   \\  
   13.57  & -     &   -   & 0.061  & 0.015  &   0.170  &  0.019 &  3.86  & 0.67  &    -  &    - &     -  &  -   \\  
   14.34  &  2.24 & 0.38  & 0.047  & 0.016  &   0.178  &  0.020 &  3.77  & 0.69  & 2.34  & 0.43 &  2.45  & 0.42 \\ 
   15.11  & -     &   -   & 0.051  & 0.017  &   0.179  &  0.021 &  3.48  & 0.73  &    -  &    - &     -  &  -   \\  
\hline
\end{tabular}
\end{table*}


\begin{table*}
\label{indices-minor.tab}
\centering
\caption{Measured line indices and their errors along the minor axis of NGC~2865.}
\vspace{0.5cm} 
\begin{tabular}{lllllllllllll}
\hline
$r (\asec)$&$H_{\beta}$&$err$&$Mg_1$&$err$&$Mg_2$&$err$& $Mg\ b$ & $err$ & $Fe_{5270}$ & $err$ & $Fe_{5335}$ & $err$ \\ \\
  -5.20 & 1.86 & 0.36 & 0.073 & 0.009 & 0.172 & 0.011 & 3.36 & 0.39 &2.21 & 0.42 & 2.26 & 0.41  \\
  -4.43 & 2.59 & 0.33 & 0.078 & 0.008 & 0.189 & 0.010 & 3.78 & 0.36 &2.92 & 0.38 & 2.11 & 0.38  \\
  -3.66 & 2.46 & 0.30 & 0.075 & 0.007 & 0.186 & 0.009 & 3.55 & 0.32 &3.20 & 0.33 & 2.48 & 0.34  \\
  -2.89 & 2.76 & 0.25 & 0.081 & 0.006 & 0.196 & 0.007 & 3.64 & 0.27 &3.11 & 0.29 & 2.98 & 0.28  \\
  -2.12 & 2.93 & 0.21 & 0.096 & 0.005 & 0.211 & 0.006 & 3.66 & 0.21 &3.21 & 0.22 & 2.74 & 0.22  \\
  -1.35 & 3.18 & 0.16 & 0.099 & 0.003 & 0.212 & 0.004 & 3.47 & 0.15 &3.23 & 0.15 & 2.86 & 0.15  \\
  -0.58 & 3.45 & 0.13 & 0.070 & 0.002 & 0.176 & 0.003 & 3.33 & 0.11 &2.68 & 0.11 & 2.80 & 0.11  \\
   0.19 & 3.49 & 0.13 & 0.043 & 0.002 & 0.162 & 0.003 & 3.43 & 0.10 &2.82 & 0.11 & 2.91 & 0.11  \\
   0.96 & 3.36 & 0.14 & 0.058 & 0.003 & 0.173 & 0.004 & 3.31 & 0.13 &2.91 & 0.14 & 2.86 & 0.14  \\
   1.73 & 3.20 & 0.17 & 0.077 & 0.004 & 0.191 & 0.005 & 3.52 & 0.19 &2.81 & 0.20 & 2.69 & 0.20  \\
   2.50 & 2.84 & 0.22 & 0.081 & 0.005 & 0.192 & 0.007 & 3.50 & 0.25 &2.90 & 0.26 & 2.79 & 0.26  \\
   3.27 & 2.53 & 0.27 & 0.066 & 0.007 & 0.193 & 0.008 & 3.89 & 0.30 &2.87 & 0.32 & 2.38 & 0.32  \\
   4.04 & 2.74 & 0.31 & 0.059 & 0.008 & 0.193 & 0.010 & 3.87 & 0.35 &2.77 & 0.37 & 2.57 & 0.37  \\
   4.81 & 2.09 & 0.36 & 0.076 & 0.009 & 0.182 & 0.011 & 3.96 & 0.39 &2.74 & 0.42 & 2.83 & 0.42  \\
   5.58 & 2.41 & 0.40 & 0.065 & 0.010 & 0.186 & 0.012 & 4.16 & 0.44 &3.32 & 0.47 & 3.53 & 0.46  \\
   6.35 & 1.93 & 0.45 & 0.073 & 0.011 & 0.196 & 0.014 & 3.27 & 0.51 &2.96 & 0.53 & 3.29 & 0.52  \\
   7.12	&   -   &  -    & 0.059 & 0.012 & 0.187 & 0.015 &  -    &  -    &  -   &  -    &   -    &-      \\
   7.89 & 2.29 & 0.31 & 0.089 & 0.013 & 0.181 & 0.016 & 4.24 & 0.33 &2.65 & 0.36 & 2.10 & 0.36  \\
   8.66	&   -   &  -    & 0.080 & 0.014 & 0.197 & 0.017 &  -    &  -    &  -   &  -    &   -    &-      \\
   9.43	&   -   &  -    & 0.072 & 0.015 & 0.140 & 0.018 &  -    &  -    &  -   &  -    &   -    &-      \\
  10.20 & 2.33 & 0.37 & 0.036 & 0.016 & 0.160 & 0.020 & 3.74 & 0.40 &2.04 & 0.44 & 1.89 & 0.44  \\
  10.97	&   -   &  -    & 0.057 & 0.016 & 0.233 & 0.021 &  -    &  -    &  -   &  -    &   -    &-      \\       
\hline
\end{tabular}
\end{table*}


\begin{thebibliography}{}

\bibitem[\protect\citeauthoryear{{Alonso}, {Arribas}, \&
  {Martinez-Roger}}{{Alonso} et~al.}{1996}]{alonso-etal96}
{Alonso} A., {Arribas} S., {Martinez-Roger} C., 1996, A\&A,{ 313}, 873

\bibitem[\protect\citeauthoryear{{Balcells}}{{Balcells}}{1992}]{Balcells92p221}
{Balcells} M., 1992, in {Longo} G., {Capaccioli} M., {Busarello} G., eds,
  Morphological and Physical Classification of Galaxies.
\newblock Kluwer Academic Publishers, Dordrecht, p.\  221

\bibitem[\protect\citeauthoryear{{Balcells}}{{Balcells}}{1997}]{balcells2tails%
97}
{Balcells} M., 1997, ApJ,{ 486}, L87

\bibitem[\protect\citeauthoryear{{Balcells} \& {Carter}}{{Balcells} \&
  {Carter}}{1993}]{BC93}
{Balcells} M., {Carter} D., 1993, A\&A,{ 279}, 376

\bibitem[\protect\citeauthoryear{{Balcells} \& {Peletier}}{{Balcells} \&
  {Peletier}}{1994}]{BP94}
{Balcells} M., {Peletier} R.F., 1994, AJ,{ 107}, 135

\bibitem[\protect\citeauthoryear{{Balcells} \& {Quinn}}{{Balcells} \&
  {Quinn}}{1990}]{BQ90}
{Balcells} M., {Quinn} P., 1990, ApJ,{ 361}, 381

\bibitem[\protect\citeauthoryear{{Barnes}}{{Barnes}}{1996}]{barnes1996fogh}
{Barnes} J.E., 1996, in {Morrison} H., {Sarajedini} A., eds, ASP Conference
  Series vol.92.
\newblock p.\  415

\bibitem[\protect\citeauthoryear{{Bender}}{{Bender}}{1990}]{bender90rev}
{Bender} R., 1990, in {Wielen} R., ed, Dynamics and Interactions of Galaxies.
\newblock Springer-Verlag, Heidelberg, p.\  232

\bibitem[\protect\citeauthoryear{{Bender}}{{Bender}}{1996}]{bender96iau}
{Bender} R., 1996, in {Bender} R., {Davies} R.L., eds, IAU Symp.~171.
\newblock Kluwer Academic Publishers, Dordrecht, p.\  181

\bibitem[\protect\citeauthoryear{{Bender}, {Saglia}, \& {Gerhard}}{{Bender}
  et~al.}{1994}]{BSG94}
{Bender} R., {Saglia} R.P., {Gerhard} O.E., 1994, MNRAS,{ 269}, 785

\bibitem[\protect\citeauthoryear{{Bender} \& {Surma}}{{Bender} \&
  {Surma}}{1992}]{BS92}
{Bender} R., {Surma} P., 1992, A\&A,{ 258}, 250

\bibitem[\protect\citeauthoryear{{Bertelli} et~al.}{{Bertelli}
  et~al.}{1994}]{bertelli-etal94}
{Bertelli} G., {Bressan} A., {Chiosi} C., {Fagotto} F., {Nasi} E., 1994,
  A\&AS,{ 106}, 275

\bibitem[\protect\citeauthoryear{{Bertschinger}}{{Bertschinger}}{1985}]{Bertsc%
hinger85}
{Bertschinger} E., 1985, ApJS,{ 58}, 39

\bibitem[\protect\citeauthoryear{{Bettoni}}{{Bettoni}}{1992}]{Bettoni92}
{Bettoni} D., 1992, A\&AS,{ 96}, 333

\bibitem[\protect\citeauthoryear{{Bica} \& {Alloin}}{{Bica} \&
  {Alloin}}{1987}]{BA87}
{Bica} E., {Alloin} D., 1987, A\&AS,{ 70}, 281

\bibitem[\protect\citeauthoryear{{Binney}}{{Binney}}{1978}]{Binney78}
{Binney} J.~J., 1978, MNRAS,{ 183}, 501

\bibitem[\protect\citeauthoryear{{Bruzual} \& {Charlot}}{{Bruzual} \&
  {Charlot}}{1997}]{BC97}
{Bruzual} G., {Charlot} S., 1997, in preparation

\bibitem[\protect\citeauthoryear{{Carollo} \& {Danziger}}{{Carollo} \&
  {Danziger}}{1994}]{carollo-danziger94}
{Carollo} M., {Danziger} I.J., 1994, MNRAS,{ 270}, 523

\bibitem[\protect\citeauthoryear{{Carter} et~al.}{{Carter}
  et~al.}{1988}]{carter-etal88}
{Carter} D., {Prieur} J-.L-., {Wilkinson} A., {Sparks} W.B., {Malin} D.F.,
  1988, MNRAS,{ 235}, 813

\bibitem[\protect\citeauthoryear{{Charlot} \& {Bruzual}}{{Charlot} \&
  {Bruzual}}{1991}]{CB91}
{Charlot} S., {Bruzual} G., 1991, ApJ,{ 367}, 126

\bibitem[\protect\citeauthoryear{{Davies}}{{Davies}}{1996}]{davies96iau}
{Davies} R.L., 1996, in {Bender} R., {Davies} R.L., eds, IAU Symp.~171.
\newblock Kluwer Academic Publishers, Dordrecht, p.\ ~37

\bibitem[\protect\citeauthoryear{{Davies} et~al.}{{Davies}
  et~al.}{1983}]{davies-etal83}
{Davies} R.L., {Efstathiou} G., {Fall} S.M., {Illingworth} G., {Schechter}
  P.L., 1983, ApJ,{ 266}, 41

\bibitem[\protect\citeauthoryear{{Davies}, {Sadler}, \& {Peletier}}{{Davies}
  et~al.}{1993}]{DSP93}
{Davies} R.L., {Sadler} E.M., {Peletier} R.F., 1993, MNRAS,{ 262}, 650

\bibitem[\protect\citeauthoryear{{De Jong} \& {Davies}}{{De Jong} \&
  {Davies}}{1997}]{dejong-davies97}
{De Jong} R.S., {Davies} R.L., 1997, MNRAS,{ 285}, L1

\bibitem[\protect\citeauthoryear{{de Vaucouleurs} et~al.}{{de Vaucouleurs}
  et~al.}{1991}]{rc3}
{de Vaucouleurs} G., {de Vaucouleurs} A., {Corwin} H.G.jr., {Buta} R.J.,
  {Paturel} G., {Fouqu\'{e}} P., 1991, Third Reference Catalog of Bright
  Galaxies.
\newblock Springer-Verlag, London


\bibitem[\protect\citeauthoryear{{Dupraz} \& {Combes}}{{Dupraz} \&
  {Combes}}{1986}]{DC86}
{Dupraz} C., {Combes} F., 1986, A\&A,{ 166}, 53

\bibitem[\protect\citeauthoryear{{Faber} et~al.}{{Faber} et~al.}{1985}]{ffbg85}
{Faber} S.M., {Friel} E.D., {Burstein} D., {Gaskell} C.M, 1985, ApJS,{ 57}, 711

\bibitem[\protect\citeauthoryear{{Fabian}, {Nulsen}, \& {Stewart}}{{Fabian}
  et~al.}{1980}]{FNS80}
{Fabian} A.C., {Nulsen} P.E.J., {Stewart} G.C., 1980, Nat,{ 287}, 613

\bibitem[\protect\citeauthoryear{{Forbes}}{{Forbes}}{1992}]{Forbes92}
{Forbes} D.A., 1992, Ph.D. Thesis, Cambridge University

\bibitem[\protect\citeauthoryear{{Fort} et~al.}{{Fort}
  et~al.}{1986}]{fort-etal86}
{Fort} B.P., {Prieur} J.-L., {Carter} D., {Meatheringham} S.J., {Vigroux} L.,
  1986, ApJ,{ 306}, 110

\bibitem[\protect\citeauthoryear{{Franx} \& {Illingworth}}{{Franx} \&
  {Illingworth}}{1988}]{fi88}
{Franx} M., {Illingworth} G.D., 1988, ApJ,{ 327}, L55

\bibitem[\protect\citeauthoryear{{Fritze-v.~Alvensleben} \&
  {Gerhard}}{{Fritze-v.~Alvensleben} \& {Gerhard}}{1994}]{fva-gerhard94}
{Fritze-v.~Alvensleben} U., {Gerhard} O.E., 1994, ApJ,{ 285}, 751

\bibitem[\protect\citeauthoryear{{Gonz\'{a}lez}}{{Gonz\'{a}lez}}{1993}]{GJ93}
{Gonz\'{a}lez} J.J., 1993, Phd thesis, Univ of California

\bibitem[\protect\citeauthoryear{{Goudfrooij} \& {Emsellem}}{{Goudfrooij} \&
  {Emsellem}}{1996}]{GE96}
{Goudfrooij} P., {Emsellem} E., 1996, A\&A,{ 306}, L45

\bibitem[\protect\citeauthoryear{{Hau}, {Balcells}, \& {Carter}}{{Hau}
  et~al.}{1996}]{hau96iau}
{Hau} G.K.T., {Balcells} M., {Carter} D., 1996, in {Bender} R., {Davies} R.L.,
  eds, IAU Symp.~171.
\newblock Kluwer Academic Publishers, Dordrecht, p.\  388

\bibitem[\protect\citeauthoryear{{Hau} \& {Thomson}}{{Hau} \&
  {Thomson}}{1994}]{HT94}
{Hau} G.K.T., {Thomson} R.C., 1994, MNRAS,{ 270}, L23

\bibitem[\protect\citeauthoryear{{Hernquist} \& {Barnes}}{{Hernquist} \&
  {Barnes}}{1991}]{HB91}
{Hernquist} L., {Barnes} J.E., 1991, Nat,{ 354}, 210

\bibitem[\protect\citeauthoryear{{Hernquist} \& {Quinn}}{{Hernquist} \&
  {Quinn}}{1987}]{HQ87}
{Hernquist} L., {Quinn} P.J., 1987, ApJ,{ 312}, 1

\bibitem[\protect\citeauthoryear{{Hernquist} \& {Quinn}}{{Hernquist} \&
  {Quinn}}{1988}]{HQ88}
{Hernquist} L., {Quinn} P.J., 1988, ApJ,{ 331}, 682

\bibitem[\protect\citeauthoryear{{Hernquist} \& {Spergel}}{{Hernquist} \&
  {Spergel}}{1992}]{HS92}
{Hernquist} L., {Spergel} D.N., 1992, ApJ,{ 399}, L117

\bibitem[\protect\citeauthoryear{{Hernquist} \& {Weil}}{{Hernquist} \&
  {Weil}}{1992}]{HW92}
{Hernquist} L., {Weil} M.L., 1992, Nat,{ 358}, 734

\bibitem[\protect\citeauthoryear{{Hibbard} \& {Mihos}}{{Hibbard} \&
  {Mihos}}{1995}]{HM95}
{Hibbard} J.E., {Mihos} J.C., 1995, AJ,{ 110}, 140

\bibitem[\protect\citeauthoryear{{Jacoby}, {Hunter}, \& {Christian}}{{Jacoby}
  et~al.}{1984}]{JHC84}
{Jacoby} G.H., {Hunter} D.A., {Christian} C.A., 1984, ApJS,{ 56}, 278

\bibitem[\protect\citeauthoryear{{J$\o$rgensen}}{{J$\o$rgensen}}{1997}]{jorgen%
sen97}
{J$\o$rgensen} I., 1997, MNRAS,{ 288}, 161

\bibitem[\protect\citeauthoryear{{J$\o$rgensen}, {Franx}, \&
  {Kjaergaard}}{{J$\o$rgensen} et~al.}{1992}]{jorgensen92}
{J$\o$rgensen} I., {Franx} M., {Kjaergaard} P., 1992, A\&AS,{ 95}, 489

\bibitem[\protect\citeauthoryear{{Keel} \& {Wu}}{{Keel} \& {Wu}}{1995}]{KW95}
{Keel} W.C., {Wu} W., 1995, AJ,{ 110}, 129

\bibitem[\protect\citeauthoryear{{Kormendy}}{{Kormendy}}{1984}]{Kormendy84}
{Kormendy} J., 1984, ApJ,{ 287}, 577

\bibitem[\protect\citeauthoryear{{L\"{o}wenstein}, {Fabian}, \&
  {Nulsen}}{{L\"{o}wenstein} et~al.}{1987}]{lowenstein-etal87}
{L\"{o}wenstein} M., {Fabian} A.C., {Nulsen} P.E.J., 1987, MNRAS,{ 229}, 129

\bibitem[\protect\citeauthoryear{{Malin} \& {Carter}}{{Malin} \&
  {Carter}}{1983}]{MC83}
{Malin} D.F., {Carter} D., 1983, ApJ,{ 274}, 534

\bibitem[\protect\citeauthoryear{{Matteucci} \& {Brocato}}{{Matteucci} \&
  {Brocato}}{1990}]{MB90}
{Matteucci} F., {Brocato} E., 1990, ApJ,{ 365}, 539

\bibitem[\protect\citeauthoryear{{Mayer} \& {Maurice}}{{Mayer} \&
  {Maurice}}{1985}]{MM85p299}
{Mayer} M., {Maurice} E., 1985, in {Davis Phillip} A.~G., {Latham} D.W., eds,
  Stellar Radial Velocities.
\newblock L. Davis Press, Schenectady, p.\  299

\bibitem[\protect\citeauthoryear{{Mehlert} et~al.}{{Mehlert}
  et~al.}{1998}]{MSBW98}
{Mehlert} D., {Saglia} R.P., {Bender} R., {Wegner} G., 1998, A\&A,{ 332}, 33

\bibitem[\protect\citeauthoryear{{Mihalas} \& {Binney}}{{Mihalas} \&
  {Binney}}{1981}]{MB81}
{Mihalas} D., {Binney} J.J., 1981, Galactic Astronomy.
\newblock Freeman, New York


\bibitem[\protect\citeauthoryear{{Mihos} \& {Hernquist}}{{Mihos} \&
  {Hernquist}}{1994}]{mihos-hernquist94apj427}
{Mihos} J.C., {Hernquist} L., 1994, ApJ,{ 427}, 112

\bibitem[\protect\citeauthoryear{{Mihos} \& {Hernquist}}{{Mihos} \&
  {Hernquist}}{1996}]{mihos-hernquist96apj464}
{Mihos} J.C., {Hernquist} L., 1996, ApJ,{ 464}, 641

\bibitem[\protect\citeauthoryear{{Quinn}}{{Quinn}}{1984}]{Quinn84}
{Quinn} P.J., 1984, ApJ,{ 279}, 596

\bibitem[\protect\citeauthoryear{{Reid}, {Boisson}, \& {Sanson}}{{Reid}
  et~al.}{1994}]{RBS94}
{Reid} N., {Boisson} C., {Sanson} A.E., 1994, MNRAS,{ 269}, 713

\bibitem[\protect\citeauthoryear{{Rix} \& {White}}{{Rix} \&
  {White}}{1992}]{RW92}
{Rix} H-.W., {White} S.D.M., 1992, MNRAS,{ 254}, 389

\bibitem[\protect\citeauthoryear{{Roberts} et~al.}{{Roberts}
  et~al.}{1991}]{roberts-etal91}
{Roberts} M.S., {Hogg} D.E., {Bregman} J.N., {Forman} W.R., {Jones} C., 1991,
  ApJS,{ 75}, 751

\bibitem[\protect\citeauthoryear{{Sandage} \& {Tammann}}{{Sandage} \&
  {Tammann}}{1991}]{rsa}
{Sandage} A., {Tammann} G., 1991, A Revised Shapley-Ames Catalog of Bright
  Galaxies.
\newblock Carnegie Inst., Washington, Washington


\bibitem[\protect\citeauthoryear{{Schiminovich} et~al.}{{Schiminovich}
  et~al.}{1995}]{schiminovich-etal95}
{Schiminovich} D., {Van Gorkom} J.H., {Van Der Hulst} J.M., {Malin} D.F., 1995,
  ApJ,{ 444}, L77

\bibitem[\protect\citeauthoryear{{Schweizer}}{{Schweizer}}{1982}]{Schweizer82}
{Schweizer} F., 1982, ApJ,{ 252}, 455

\bibitem[\protect\citeauthoryear{{Scorza} \& {Bender}}{{Scorza} \&
  {Bender}}{1990}]{scorza-bender90}
{Scorza} C., {Bender} R., 1990, A\&A,{ 235}, 49

\bibitem[\protect\citeauthoryear{{Thomson}}{{Thomson}}{1991}]{Thomson91}
{Thomson} R.C., 1991, MNRAS,{ 253}, 256

\bibitem[\protect\citeauthoryear{{Thomson} \& {Wright}}{{Thomson} \&
  {Wright}}{1990}]{TW90}
{Thomson} R.C., {Wright} A.E., 1990, MNRAS,{ 247}, 122

\bibitem[\protect\citeauthoryear{{Trager}}{{Trager}}{1997}]{trager-etal97}
{Trager} et.~al., 1997, in preparation

\bibitem[\protect\citeauthoryear{{van der Marel} \& {Franx}}{{van der Marel} \&
  {Franx}}{1993}]{vdmf93}
{van der Marel} R.P., {Franx} M., 1993, ApJ,{ 407}, 525

\bibitem[\protect\citeauthoryear{{van der Marel} et~al.}{{van der Marel}
  et~al.}{1994}]{vdm-etal94}
{van der Marel} R.P., {Rix} H.-W., {Carter} D., {Franx} M., {White} S.D.M., {De
  Zeeuw} T., 1994, MNRAS,{ 268}, 521

\bibitem[\protect\citeauthoryear{{Vila-Costas} \& {Edmunds}}{{Vila-Costas} \&
  {Edmunds}}{1992}]{VE92}
{Vila-Costas} M.~B., {Edmunds} M., 1992, MNRAS,{ 259}, 121

\bibitem[\protect\citeauthoryear{{Williams} \& {Christiansen}}{{Williams} \&
  {Christiansen}}{1986}]{WC86}
{Williams} R.E., {Christiansen} W.A., 1986, ApJ,{ 291}, 80

\bibitem[\protect\citeauthoryear{{Worthey} et~al.}{{Worthey}
  et~al.}{1994}]{worthey-etal94apjs94}
{Worthey} G., {Faber} S.M., {Jes\'{u}s Gonz\'{a}lez} J., {Burstein} D., 1994,
  ApJS,{ 94}, 687

\end{thebibliography}
\end{document}